\pdfoutput=1

\documentclass[a4paper,fleqn,usenatbib]{mnras}

\usepackage{newtxtext,newtxmath}

\usepackage[T1]{fontenc}
\usepackage{ae,aecompl}

\usepackage[usenames]{color}

\newcommand{\dsct}{$\delta$~Sct }
\newcommand{\gdor}{$\gamma$~Dor }
\newcommand{\Kepler}{\textit{Kepler} }

\usepackage{graphicx}	
\usepackage{amsmath}	
\usepackage{amssymb}	

\title[Characterising \Kepler \dsct stars]{Characterising the observational properties of \dsct stars in the era of space photometry from the \Kepler mission}

\author[D. M. Bowman and D. W. Kurtz]{
Dominic M. Bowman$^{1,2}$\thanks{E-mail: dominic.bowman@kuleuven.be (DMB)} and Donald W. Kurtz$^{2}$
\\
$^{1}$ Instituut voor Sterrenkunde, KU Leuven, Celestijnenlaan 200D, 3001 Leuven, Belgium \\
$^{2}$ Jeremiah Horrocks Institute, University of Central Lancashire, Preston PR1 2HE, UK \\
}

\date{Accepted 2018 February 14. Received 2018 February 13; in original form 2017 December 22}

\pubyear{2018}

\begin{document}
\label{firstpage}
\pagerange{\pageref{firstpage}--\pageref{lastpage}}
\maketitle

\begin{abstract}
The \dsct stars are a diverse group of intermediate-mass pulsating stars located on and near the main sequence within the classical instability strip in the Hertzsprung--Russell diagram. Many of these stars are hybrid stars pulsating simultaneously with pressure and gravity modes that probe the physics at different depths within a star's interior. Using two large ensembles of \dsct stars observed by the \Kepler Space Telescope, the instrumental biases inherent to \Kepler mission data and the statistical properties of these stars are investigated. An important focus of this work is an analysis of the relationships between the pulsational and stellar parameters, and their distribution within the classical instability strip. It is found that a non-negligible fraction of main sequence \dsct stars exist outside theoretical predictions of the classical instability boundaries, which indicates the necessity of a mass-dependent mixing length parameter to simultaneously explain low- and high-radial order pressure modes in \dsct stars within the Hertzsprung--Russell diagram. Furthermore, a search for regularities in the amplitude spectra of these stars is also presented, specifically the frequency difference between pressure~modes of consecutive radial order. In this work, it is demonstrated that an ensemble-based approach using space photometry from the \Kepler mission is not only plausible for \dsct stars, but that it is a valuable method for identifying the most promising stars for mode identification and asteroseismic modelling. The full scientific potential of studying \dsct stars is as yet unrealised. The ensembles discussed in this paper represent a high-quality data set for future studies of rotation and angular momentum transport inside A and F stars using asteroseismology.
\end{abstract}




\section{Introduction}
\label{intro}

The \dsct stars are a diverse group of pulsating stars found at the intersection of the main sequence and the classical instability strip in the Hertzsprung--Russell (HR) diagram, which corresponds to spectral types between A2\,V and F2\,V for main sequence Population~{\sc I} stars, and effective temperatures approximately between $6400 \leq T_{\rm eff} \leq 8600$ \citep{Breger2000b, Rod2001, ASTERO_BOOK, Uytterhoeven2011}. This places \dsct stars within an interesting transition region in the HR~diagram between low-mass stars with radiative cores and thick convective envelopes ($M \lesssim 1$~M$_{\rm \odot}$) and high-mass stars with large convective cores and radiative envelopes ($M \gtrsim 2$~M$_{\rm \odot}$) --- see \citet{Bowman_BOOK} for a recent review of \dsct stars. The observations of the physical properties of \dsct stars provide useful constraints for stellar structure and evolutionary models, since the physical properties such as interior rotation, and the size and shape of the convective core have a significant impact on the evolution of intermediate and massive stars \citep{Maeder_rotation_BOOK, Meynet_rotation_BOOK}.

The observation and modelling of stellar pulsations -- known as asteroseismology -- provides insight of the interior physics of stars for a wide range of stellar masses and evolutionary stages across the HR~diagram \citep{ASTERO_BOOK}. The pulsation modes excited within a star are dependent on stellar structure, thus pulsation mode frequencies provide direct insight of physics within a star where traditional methods in astronomy, such as photometry and spectroscopy, are unable to probe directly. This makes asteroseismology unique, with its strong observational constraints of stellar physics used for improving theoretical models of stellar structure and evolution. The dominant pulsations in \dsct stars are self excited by the opacity ($\kappa$) mechanism operating in the He~{\sc ii} ionisation zone \citep{Cox1963, Breger2000b, ASTERO_BOOK}, which produces pressure (p) modes with periods as long as eight hours and short as 15~min \citep{Uytterhoeven2011, Holdsworth2014c}. These p~modes typically have low radial orders and are most sensitive to the surface layers of a star. Another form of coherent mode excitation in \dsct stars is from turbulent pressure, which is able to excite p~modes with radial orders between $7 \leq n \leq 10$ with higher pulsation mode frequencies for stars within the classical instability strip \citep{Houdek2000, Antoci2014b, Xiong2016a}. 

The instability regions of the \dsct and the lower-mass \gdor stars are predicted to overlap in the HR~diagram \citep{Dupret2004, Dupret2005, Houdek2015}, with stars that simultaneously pulsate in p~modes excited by the $\kappa$~mechanism and gravity (g) modes excited by the convective flux blocking (modulation) mechanism commonly referred to as hybrid stars. The exact nature the excitation mechanism of g~modes and its physical relationship with the $\kappa$-mechanism in intermediate-mass stars is not fully understood (see e.g., \citealt{Xiong2016a}), with a significant fraction of \dsct stars hotter than the \gdor instability region observed to pulsate with g~modes \citep{Uytterhoeven2011, Bowman_PhD, Bowman_BOOK}. Few hybrid stars were known prior to space photometry (see e.g., \citealt{Handler2009c}). The unprecedented photometric precision, duty cycle and total length of the \Kepler mission \citep{Borucki2010, Koch2010} revealed that many \dsct stars are hybrid stars \citep{Griga2010a, Uytterhoeven2011, Balona2011g, Balona2014a}. On the other hand, it is also demonstrably true that pure \dsct stars with no significant frequencies below 5~d$^{-1}$, albeit rare, exist \citep{Bowman_BOOK}. 

An analysis of 750 A and F stars was carried out by \citet{Uytterhoeven2011} using 1~yr (Q0 -- Q4) of \Kepler data, and it was concluded that approximately 25~per~cent were hybrid pulsators. Later analyses of A and F stars using longer time spans of \Kepler data by \citet{Balona2014a} and \citet{Balona2015e} have found much higher fractions of \dsct stars with low frequencies in their amplitude spectra, but also demonstrate that less than fifty~per~cent of stars within the classical instability strip pulsate. This measured incidence of pulsation amongst \dsct stars should be considered a lower limit with 50~per~cent of A and F stars observed to pulsate using the detection threshold \Kepler photometry, which is of order a few $\mu$mag in the best of cases. For many A and F pulsators, it is not established if observed low-frequency peaks are caused by pulsation modes, rotational modulation, combination frequencies or have some other astrophysical cause \citep{Bowman_BOOK}. Recent studies by \citet{VanReeth2016a} and \citet{Saio2018a} have discovered that global normal modes of Rossby waves (r modes), which cause temperature perturbations in a rotating star, are visible in A and F stars in the \Kepler data set. Specifically, r modes in rapidly rotating \gdor stars exhibit a power excess at low frequency, such that r modes with azimuthal order $m$ produce a group of frequencies slightly below $m$ times a star's rotation frequency \citep{Saio2018a}. 

For chemically normal stars within the classical instability strip, it is not surprising that they are observed to pulsate from the nature of the driving mechanism in \dsct stars \citep{Murphy2015a}, although it is also known that approximately half of stars within the classical instability strip are not pulsating \citep{Balona2011g} at the photometric precision of order a few $\mu$mag in \Kepler data. Further work is needed to study the interplay between the flux blocking mechanism and the $\kappa$~mechanism in A and F stars, with a focus on understanding the synergy between the pulsation excitation mechanisms and the observed minority of \dsct stars pulsating in purely p modes. This provides motivation for characterising the pulsation properties of a large ensemble of \dsct stars observed by \Kepler presented in this work.

However, one of the main difficulties encountered when observing \dsct stars is the issue of identifying pulsation modes in terms of their radial order, $n$, angular degree, $\ell$, and azimuthal order, $m$. The typical low-order p modes in \dsct stars often do not show any regular spacing in the amplitude spectrum, unlike the asymptotic high radial order p modes in low-mass solar-type stars --- see \citet{Chaplin2013c} and \citet{Hekker2017a} for reviews of solar-like pulsators. The rapid rotation, mode visibility and our incomplete understanding of mode excitation in \dsct stars make them one of the more challenging groups of pulsating stars to study using asteroseismology (e.g., \citealt{Reese2017a}). Despite these complexities, ongoing efforts to apply ensemble asteroseismology techniques to \dsct stars are being made using similar methods to those used to study low-mass stars \citep{GH2009, GH2013, GH2015, Paparo2016b, Michel2017b}.

	Significant progress in understanding the physics of \dsct stars has been made since the space photometry revolution facilitated by the CoRoT \citep{Auvergne2009} and \Kepler \citep{Borucki2010} space missions. The rich pulsation spectra of \dsct stars offer the potential to probe physics at different depths using asteroseismology, thus these stars can provide useful constraints of stellar structure and evolution theory in an important transition region on the main sequence \citep{Breger2000b, ASTERO_BOOK}. The full scientific potential of studying pulsation modes in \dsct stars with asteroseismology in the era of space photometry has yet to be exploited.


	\subsection{Regularities in the amplitude spectra of \dsct stars}
	\label{subsection: regular}
		
	It has been the goal of various studies to search for regularities in the amplitude spectra of \dsct stars (e.g., \citealt{GH2009, GH2013, GH2015, Paparo2016b, Michel2017b}), with a particular focus of identifying the asymptotic frequency separation between modes of consecutive high radial order, known as the large frequency separation $\Delta\nu$. However, \dsct stars often have high mode densities in their amplitude spectra caused by dozens, and sometimes hundreds, of pulsation mode frequencies making it difficult to distinguish intrinsic pulsation modes. This is complicated by non-linearity of pulsation modes in \dsct stars in the form of harmonics, $n\nu_i$, and combination frequencies, $n\nu_i \pm m\nu_j$, where $n$ and $m$ take integer values \citep{Papics2012b, Kurtz2015b, Bowman_BOOK}. These non-linear combination frequencies can have high Fourier amplitudes but do not represent intrinsic pulsation modes since they describe the non-sinusoidal shape of a star's light curve in the Fourier domain, and can also create regularities in an amplitude spectrum (see chapter~6 from \citealt{Bowman_BOOK}).
	
	Recently, \citet{Michel2017b} searched for regularity in the amplitude spectra of 1860 \dsct stars observed by CoRoT, using two observables which they termed $f_{\rm min}$ and $f_{\rm max}$ corresponding to the lower and upper limits for significant frequencies in a star's amplitude spectrum. \citet{Michel2017b} define significant frequencies as those that have an amplitude larger than ten times the mean amplitude in an amplitude spectrum. A subsample of approximately 250 of these stars were classified as young \dsct stars since they had high-frequency pulsation modes with $f_{\rm max} \geq 400~\mu$Hz $(\gtrsim 35$~d$^{-1})$, such that they were consistent with theoretical modes of stars near the zero-age main sequence (ZAMS). \citet{Michel2017b} found an agreement between the observed frequency distribution and the corresponding distribution caused by island modes from theoretical models using a non-perturbative treatment of fast rotation \citep{Reese2009a}. Specifically, regularities in the amplitude spectra of these stars produced ridge-like structures with spacings of order a few tens of $\mu$Hz (of order a few d$^{-1}$) consistent with consecutive radial order pulsation modes. Thus, the observables $f_{\rm min}$ and $f_{\rm max}$ were concluded to be potentially useful diagnostics to constrain the mass and evolutionary stage of a \dsct star \citep{Michel2017b}.
	
	The ensemble approach of studying pulsating stars certainly has its advantages, especially for such diverse pulsators as \dsct stars \citep{Griga2010a, Uytterhoeven2011, Balona2011g, Balona2014a, Bowman2016a, Bowman_BOOK, Michel2017b}. In this paper, the statistical properties of a large number of \dsct stars observed by the \Kepler Space Telescope are investigated with a discussion of the target selection methods provided in section~\ref{section: ensemble}, an evaluation of the edges of the classical instability strip presented in section~\ref{section: Instability Strip}, an analysis of the correlations between the pulsational and stellar parameters given in section~\ref{section: characterising}, and the results for a search for regularities in the amplitude spectra given in section~\ref{section: regularities}. Finally, the conclusions are presented in section~\ref{section: conclusions}.


\section{Compiling ensembles of \Kepler \dsct stars}
\label{section: ensemble}

The \Kepler Space Telescope was launched on 2009~March~7 and has an Earth-trailing orbit with a period of 372.5~d \citep{Borucki2010}. Within its 115~deg$^2$ field of view in the constellations of Cygnus and Lyra, it observed approximately 200\,000 stars at photometric precision of order a few $\mu$mag \citep{Koch2010}. To maximise its goal of finding transiting Earth-like planets orbiting Sun-like stars, \Kepler had two observing modes: a long cadence (LC) of 29.45~min and a short cadence (SC) of 58.5~s \citep{Gilliland2010}. From the limited onboard data storage capacity, only 512 stars could be observed in SC at a given epoch. The \Kepler LC data were divided into quarters (from Q0 -- Q17) covering a maximum length of 1470.5~d, whereas SC data were divided into 30-d segments, which was motivated by the rolling of the \Kepler spacecraft and data downlink epochs approximately every 90 and 30~d, respectively. \Kepler data are available from the Mikulski Archive for Space Telescopes (MAST\footnote{MAST website: \url{http://archive.stsci.edu/kepler/}}) in pre- and post-pipeline formats --- see \citet{Smith2012} and \citet{Stumpe2012} for details of the \Kepler science data pipeline. In this work, the post-pipeline data, known as msMAP PDC data, are used.

Previously, a large ensemble of 983 \dsct stars with LC \Kepler data was compiled by \citet{Bowman2016a} to study amplitude modulation of pulsation modes. Later, \citet{Bowman_PhD, Bowman_BOOK} used the same stars to investigate the relationship between the pulsation and stellar parameters in \dsct stars. This ensemble of \dsct stars was comprised of \Kepler stars characterised by $6400 \leq T_{\rm eff} \leq 10\,000$~K in the \Kepler Input Catalogue (KIC; \citealt{Brown2011}); were observed continuously in LC by \Kepler for 4~yr; and had significant peaks in the p~mode frequency range ($\nu \geq 4$~d$^{-1}$) with amplitudes above 0.10~mmag. Although an amplitude cut-off of 0.10~mmag is higher than the typical noise level observed for \Kepler data, it was chosen by \citet{Bowman2016a} to ensure that all extracted peaks had reasonable phase uncertainties, as these are dependent on the amplitude signal-to-noise ($S/N$) ratio \citep{Montgomery1999}. The KIC ID numbers and stellar parameters such as effective temperature and surface gravity of these stars were provided as supplementary online data in \citet{Bowman2016a}, with a machine-readable table also available through CDS.

However, as discussed by \citet{Bowman_PhD, Bowman_BOOK}, this ensemble of \dsct stars using LC \Kepler data suffers from instrumental biases. Firstly, the \Kepler wavelength passband of $420 - 900$~nm \citep{Koch2010} was chosen to maximise the goal of finding transiting exoplanets around Sun-like stars. This leads to the pulsation mode amplitudes of the early-type stars observed by the \Kepler mission, including \dsct stars, having lower pulsation mode amplitudes when observed in the white-light \Kepler passband compared to bluer passbands such as Johnson~$B$ \citep{Bowman2015a, Holdsworth2018b*}. Furthermore, the amplitude visibility function (which is termed {\it apodization} by \citealt{Hekker2017a}) is a strong function of frequency, such that pulsation mode amplitudes are heavily suppressed near integer multiples of the instrumental sampling frequency, which corresponds to 48.9~d$^{-1}$ and 1476.9~d$^{-1}$ for LC and SC \Kepler data, respectively \citep{Bowman_PhD, Bowman_BOOK}. This amplitude visibility function is given by:
\begin{equation}
A = A_0\,{\rm sinc}\left( \frac{\pi}{n} \right) = A_0\,{\rm sinc}\left( \frac{\pi \nu}{\nu_{\rm samp}} \right) ~ ,
\label{equation: amp visibility}
\end{equation}
\noindent where $A$ is the observed amplitude, $A_0$ is the true amplitude, $n$ is the number of data points per pulsation cycle, $\nu$ is the pulsation mode frequency and $\nu_{\rm samp}$ is the instrumental sampling frequency. Thus, for example a hypothetical \dsct star that has a pulsation mode frequency of $\nu \simeq 40$~d$^{-1}$ and an intrinsic amplitude of 1~mmag would be observed with an amplitude of 0.2~mmag using LC \Kepler data. This amplitude suppression represents a noticeable bias towards extracting low-frequency pulsation modes over high-frequency pulsation modes in any iterative pre-whitening procedure, and by extension, a bias towards studying cool and/or evolved \dsct stars when using LC \Kepler mission data \citep{Bowman_PhD, Bowman_BOOK}. These selection effects are significant when comparing results obtained with LC and SC \Kepler data, especially for asteroseismology of \dsct stars as they pulsate with mode frequencies that can span the \Kepler LC sampling frequency. 

To evaluate the impact of these biases in the LC \Kepler data for the analysis of \dsct stars, specifically the amplitude visibility function suppressing high-frequency pulsation modes, a second ensemble of \dsct stars compiled using SC \Kepler data is also analysed independently in this work. The \dsct stars in the SC ensemble of stars were selected using the same criteria as the LC ensemble, except no restriction was placed on the length of the SC data. The ensembles of SC and LC \dsct stars have been filtered for known binarity using the Villanova binary catalogue\footnote{an updated list of known \Kepler eclipsing binary stars can be found at: \url{http://keplerebs.villanova.edu}} \citep{Prsa2011, Abdul-Masih2016a}. It is important to note that \dsct stars in eclipsing binaries and ellipsoidal variables have been excluded from the ensembles presented in this work because of the effects of contamination, which could significantly affect the $T_{\rm eff}$ determination using photometry such as those included in the KIC \citep{Brown2011}.

These selection criteria produced a LC ensemble of {963} stars and a SC ensemble of {334} \dsct stars. All stars in the LC ensemble have continuous \Kepler observations spanning 4~yr, but the \dsct stars in the SC ensemble have total time spans that range between 10 and 1470~d. The longest continuous light curve for each SC \dsct star ranges between 10 and 1000~d, with approximately 50~per~cent of stars having only 30~d of \Kepler SC data. There are {218} stars that are duplicated in both ensembles, with {116} SC stars that are not included in the LC ensemble. It is easy to understand why additional SC \dsct stars were not included in the LC ensemble, in terms of amplitude suppression from the amplitude visibility function given in Eqn.~(\ref{equation: amp visibility}), and because SC observing slots were at a premium during the \Kepler mission. The duplicate stars are not removed from either ensemble, nor are the ensembles combined as each ensemble suffers from different biases which should not be compounded into unknown biases.


	\subsection{Extracting \dsct stellar parameters}
	\label{subsection: stellar parameters}
	
	The vast majority of the 200\,000 \Kepler target stars were characterised with values of $T_{\rm eff}$ and $\log\,g$ using $griz$ and 2MASS $JHK$ broad-band photometry in the KIC \citep{Brown2011}. This catalogue has since been revised by \citet{Huber2014} with values of $T_{\rm eff}$ being accurate in a statistical sense for such a large number of stars.

	High-resolution and high $S/N$ spectroscopy using the {\sc HERMES} spectrograph mounted on the Mercator telescope \citep{Raskin2011} was obtained and analysed by \citet{Niemczura2015} for 117 bright ($V \lesssim 10$~mag) A and F stars observed by {\it Kepler}. Accurate values for fundamental stellar parameters including $T_{\rm eff}$, $\log\,g$, rotational and micro-turbulent velocities and their respective uncertainties were determined for each star \citep{Niemczura2015}. An additional 44 A and F stars were also observed using the {\sc FIES} spectrograph mounted on the Nordic Optical Telescope by \citet{Niemczura2017a}. Similar to the analysis by \citet{Huber2014}, a $200$~K systematic offset was found between the KIC and spectroscopic values of $T_{\rm eff}$ \citep{Niemczura2015, Niemczura2017a} for A and F stars. High-resolution and high $S/N$ spectroscopy of B, A and F stars observed by \Kepler was also obtained by \citet{Tkachenko2012a} and \citet{Tkachenko2013a} using the 2-m telescope of the Th{\" u}ringer Landessternwarte Tautenburg in Germany. 
	
	However, not all the stars studied by \citet{Niemczura2015, Niemczura2017a} and \citet{Tkachenko2012a, Tkachenko2013a} are \dsct stars, but those that are represent a subset of \dsct stars for which accurate values of $T_{\rm eff}$ and $\log\,g$ are known, with uncertainties for an individual star typically between $100 \leq \sigma\left(T_{\rm eff}\right) \leq 200$~K and $0.1 \leq \sigma\left(\log\,g\right) \leq 0.2$~(cgs), respectively. Only {58} and {70} stars studied by \citet{Tkachenko2012a, Tkachenko2013a} and \citet{Niemczura2015, Niemczura2017a} are included in the LC and SC ensembles, respectively, but these stars are included in this work to show compatibility with the KIC and \citet{Huber2014} values.


	\subsection{Extracting \dsct pulsation parameters}
	\label{subsection: pulsation parameters}
	
	To characterise the pulsational properties for the \dsct stars in both ensembles, the amplitude spectrum for each star was calculated from $0 \leq \nu \leq 98$~d$^{-1}$, which has an upper frequency limit corresponding to four times the LC Nyquist frequency. The highest amplitude pulsation mode was extracted and optimised using a non-linear least-squares fit to the light curve using the function $\Delta\,m = A\,\cos(2\pi\nu(t -t_0)+\phi)$, where $A$, $\nu$ and $\phi$ are the pulsation mode amplitude, frequency and phase, respectively, calculated using the zero-point of the time scale as $t_0$ = 2\,455\,688.77~BJD. Each extracted pulsation mode frequency was checked to satisfy the commonly-used significance criterion of $S/N \geq 4$ in amplitude \citep{Breger1993b} and that it did not correspond to a known instrumental spurious frequency\footnote{\Kepler data release notes: \url{https://archive.stsci.edu/kepler/data_release.html}}. The amplitude spectrum over a large frequency range for each star is useful for studying regularities in \dsct stars, which is discussed later in section~\ref{section: regularities}.	
	
	By not correcting the amplitude spectra for the \dsct stars in either ensemble for the amplitude visibility function given in Eqn.~(\ref{equation: amp visibility}) prior to extracting pulsation mode frequencies, it is ensured that the extracted peak is a pulsation mode frequency and not a Nyquist alias frequency, since Nyquist aliases have lower observed amplitudes than the corresponding real peaks \citep{Murphy2013a}. Furthermore, this approach is to avoid dividing by zero for pulsation mode frequencies very close to the LC \Kepler sampling frequency, which would cause some \dsct stars to have extremely large and unphysical pulsation mode amplitudes. Instead, pulsation mode amplitudes are corrected for the amplitude visibility function after they have been extracted from an amplitude spectrum. The disadvantage of this approach is that the distribution of LC pulsation mode frequencies will be biased towards, on average, lower values, but this is a compromise that yields the most meaningful results. This also emphasises the need to keep both ensembles of \dsct stars separate in this analysis, since the amplitude visibility function has a negligible effect for stars in the SC \Kepler ensemble.


\section{Revisiting the edges of the Classical Instability Strip with \Kepler data}
\label{section: Instability Strip}

In this section, the observed boundaries of the classical instability strip are evaluated using the high-quality photometry provided by the \Kepler Space Telescope using the two ensembles of \dsct stars. Prior to space photometry from CoRoT and {\it Kepler}, evolutionary and pulsation models of \dsct stars were calibrated to catalogues of \dsct stars collated using ground-based observations, for example, \citet{Rod2001}. Compared to space data, ground-based observations have higher noise levels and suffer from lower duty cycles and significant aliasing, which introduces strong selection effects when studying pulsating stars. These issues make it difficult to extract low-amplitude pulsation modes and thus a \dsct star may be classified as a non-pulsating star from the ground if its pulsation mode amplitudes are lower than the noise level. Typically, the noise level in the amplitude spectrum of ground-based photometry is a few tenths of a mmag for a single observing site (e.g., \citealt{Bowman2015a, Holdsworth2018b*}, but can be as low as tens of $\mu$mag in the best of circumstances \citep{Kurtz2005b}. With \Kepler data, we are afforded the luxury of 4~yr of continuous high-precision light curves resulting in a white-noise level of order a few $\mu$mag in an amplitude spectrum.

\begin{figure*}
\centering
\includegraphics[width=\columnwidth]{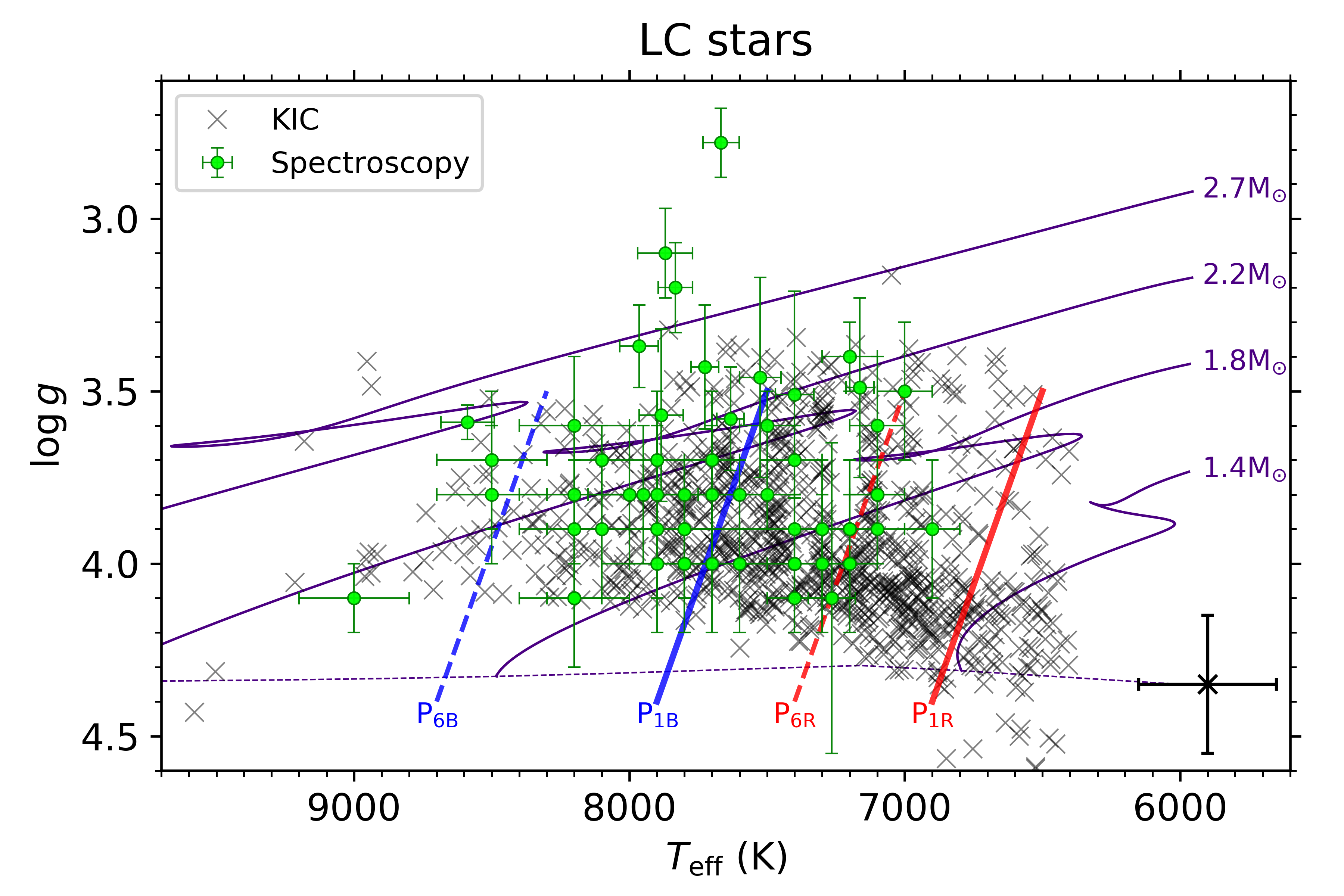}
\includegraphics[width=\columnwidth]{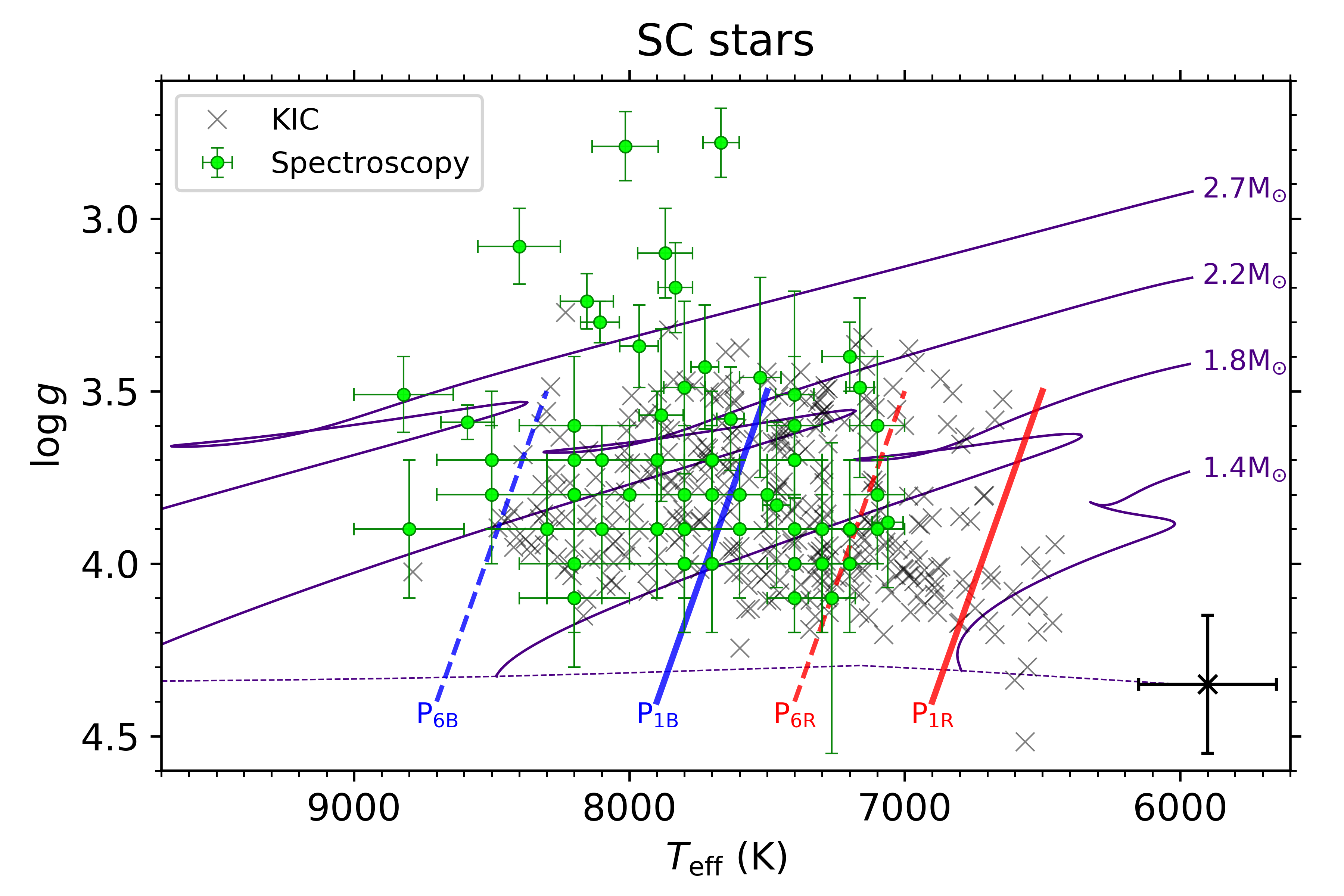}
\includegraphics[width=\columnwidth]{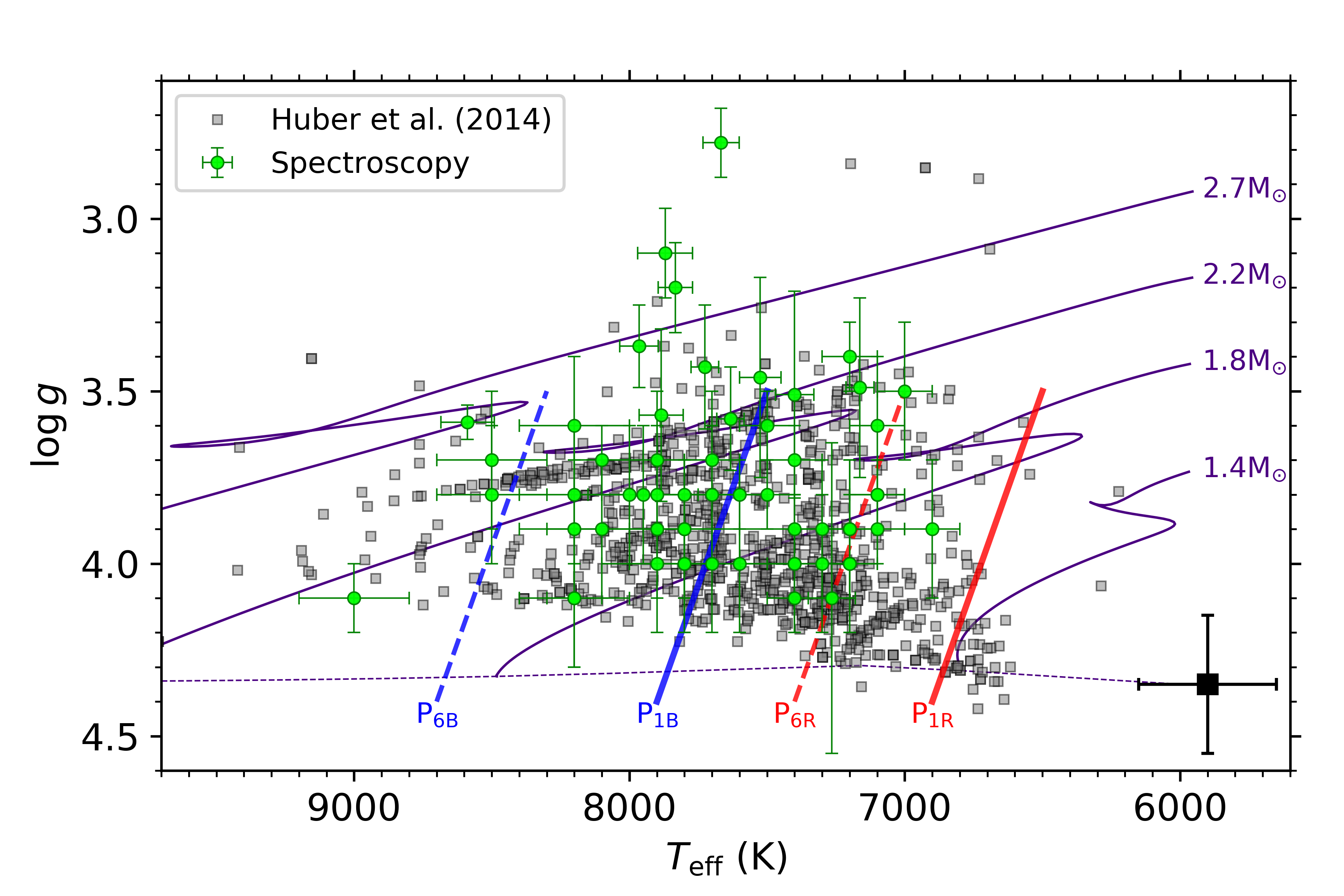}
\includegraphics[width=\columnwidth]{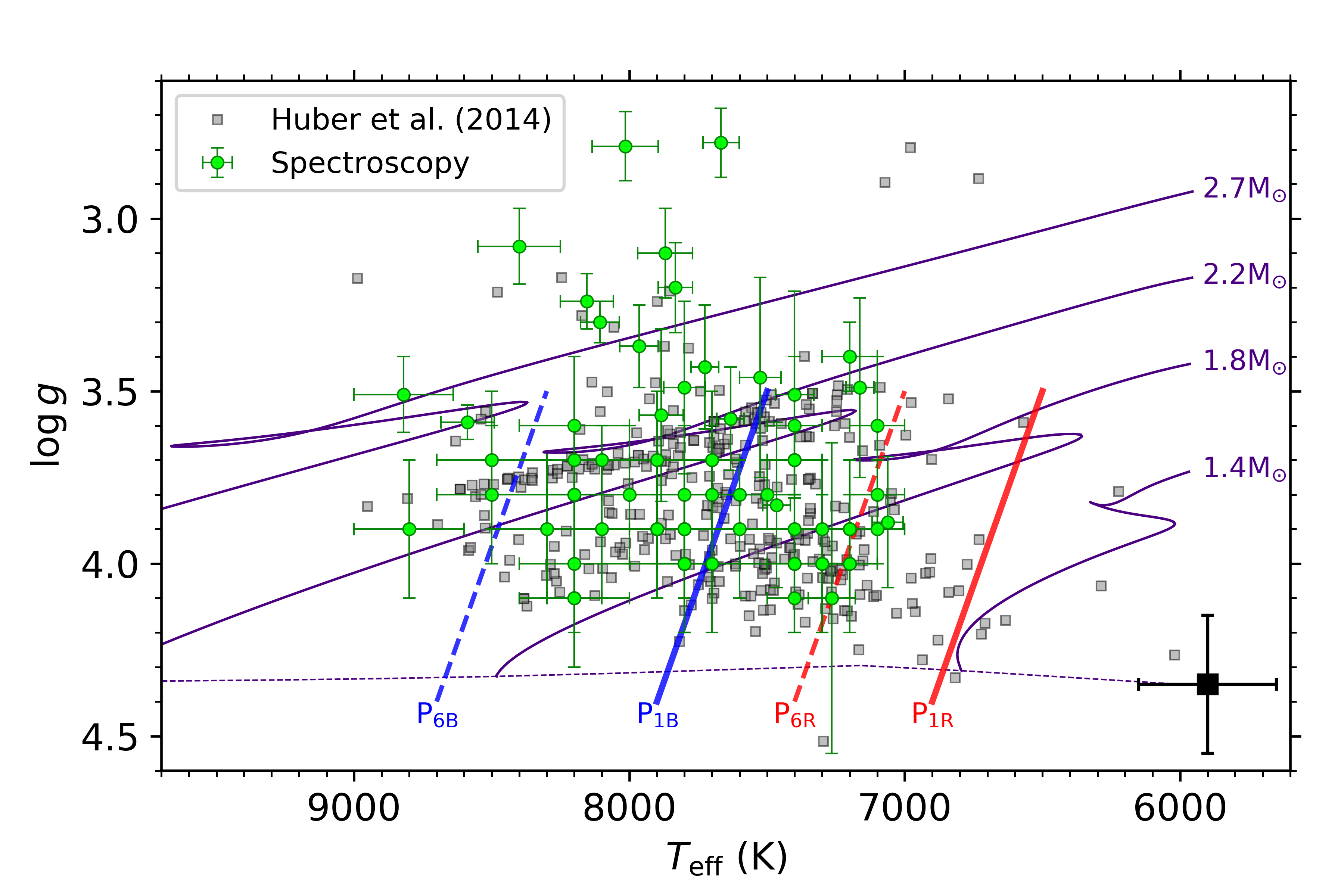}
\caption{The $T_{\rm eff} - \log\,g$ diagrams for the ensemble of LC and SC \dsct stars observed by \Kepler are shown in the {left} and {right} panels, respectively. The locations of \dsct stars using parameters from the KIC are shown by black crosses in the top row, and \citet{Huber2014} values are shown as grey squares in the bottom row, with a typical error bar shown in bold in the bottom-right corner of each panel. Stellar evolutionary tracks using a metallicity of $Z = 0.02$ from \citet{Griga2005} calculated between the ZAMS (shown as a short-dashed purple line) and a post main sequence effective temperature of $T_{\rm eff} = 5900$~K are shown as solid purple lines. The subgroup of \dsct stars with spectroscopic values of $T_{\rm eff}$ and $\log\,g$ from \citet{Tkachenko2012a, Tkachenko2013a} and \citet{Niemczura2015, Niemczura2017a} are plotted as filled green circles. The theoretical blue and red edges of the classical instability strip from \citet{Dupret2005} for p~modes with radial order of $n=1$ are shown as solid blue and red lines, respectively, and radial order of $n=6$ shown as dashed blue and red lines, respectively. }
\label{figure: Teff-logg}
\end{figure*}

The distribution of the {963} LC and {334} SC \dsct stars in $T_{\rm eff}-\log\,g$ diagrams using values from the KIC are shown in the {top-left} and {top-right} panels of Fig.~\ref{figure: Teff-logg}, respectively. The location of each \dsct star is shown by a black cross, with a typical error bar shown in bold in the bottom-right corner of each panel. For comparison, the same ensembles using the revised parameters from \citet{Huber2014} are shown in bottom-left and bottom-right panels of Fig.~\ref{figure: Teff-logg} as filled grey squares. Stellar evolutionary tracks using a metallicity of $Z = 0.02$ from \citet{Griga2005}, which are calculated from the ZAMS (indicated by a short-dashed purple line) to a post-MS effective temperature of $T_{\rm eff} = 5900$~K, for stars of 1.4, 1.8, 2.2 and 2.7~M$_{\rm \odot}$ are also shown in Fig.~\ref{figure: Teff-logg} as solid purple lines. The theoretical blue and red edges of the classical instability strip from \citet{Dupret2005} for radial p~modes between $1 \leq n \leq 6$ are shown as blue and red lines, with solid and dashed lines indicating the boundaries for $n=1$ and $n=6$ radial modes, respectively. The subgroup of \dsct stars for which accurate spectroscopic values of $T_{\rm eff}$ and $\log\,g$ from \citet{Tkachenko2012a, Tkachenko2013a} and \citet{Niemczura2015, Niemczura2017a} are also plotted as filled green circles in each panel with their respective 1-$\sigma$ uncertainties in Fig.~\ref{figure: Teff-logg}. 

The majority of \dsct stars shown in Fig.~\ref{figure: Teff-logg} are schematically consistent with the theoretical blue and red edges of the classical instability strip calculated by \citet{Dupret2005} if the lower and upper bounds of the uncertainties on $T_{\rm eff}$ values are considered for hot and cool \dsct stars, respectively. However, there is no evidence to indicate that temperatures of hot and cool \dsct stars would be over- and under-estimated in such a systematic fashion. When comparing the distribution of stars using the KIC and \citet{Huber2014} values in Fig.~\ref{figure: Teff-logg}, it is clear that the \citet{Huber2014} values are shifted to temperatures approximately 200~K higher. However, this systematic increase in temperature of the \citet{Huber2014} compared to the KIC values produces more outliers beyond the blue edge. These outliers represent a small yet significant fraction of \dsct stars in both ensembles, which includes stars with accurate spectroscopic $T_{\rm eff}$ values obtained by \citet{Tkachenko2012a, Tkachenko2013a} and \citet{Niemczura2015, Niemczura2017a}, that are significantly hotter than the blue edge for the $n=6$ radial overtone p~mode calculated by \citet{Dupret2005}. 

Many of the \dsct stars in Fig.~\ref{figure: Teff-logg} have surface gravities between $3.5 \leq \log\,g \leq 4.0$, except for those near to the ZAMS and the red edge. Thus the middle of the main sequence and the Terminal Age Main Sequence (TAMS) are well sampled compared to the ZAMS for hot \dsct stars observed by the \Kepler Space Telescope. The high density of \dsct stars shown in the left panels of Fig.~\ref{figure: Teff-logg} that are near the ZAMS and red edge of the theoretical instability strip  calculated by \citet{Dupret2005}, implies that ZAMS stars cooler than $T_{\rm eff} \simeq 6900$~K are able to pulsate in p~modes. Many of the outliers at the red edge of the classical instability strip can be explained as within $1\sigma$ of the classical instability strip, but equally importantly, many of these \dsct stars are within $1\sigma$ of being cooler than the red edge. The lack of \dsct stars with spectroscopic effective temperatures and surface gravities from \citet{Tkachenko2012a, Tkachenko2013a} and \citet{Niemczura2015, Niemczura2017a} that are cooler than the red edge is likely caused by a selection effect in these author's target lists.

A value of $\alpha_{\rm MLT}$ between 1.8 and 2.0 was determined to produce a reasonably good agreement between the theoretical models and observations of \dsct stars by \citet{Dupret2004, Dupret2005}. However, it should be noted that the highest mass model used by \citet{Dupret2005} to investigate $\alpha_{\rm MLT}$ was 1.8~M$_{\rm \odot}$, but higher-mass \dsct stars exist within the \Kepler data set, especially since stars with masses $M > 2.7$~M$_{\rm \odot}$ are included in Fig.~\ref{figure: Teff-logg} and have been confirmed by spectroscopy. It was discussed in detail by \citet{Dupret2005} that decreasing the value of $\alpha_{\rm MLT}$ to 1.0 has the effect of shifting the instability strip to lower values of effective temperature, and vice versa. Thus, a mass-dependent value of $\alpha_{\rm MLT}$ is needed to reproduce both the {\it observed} blue and red edges of the classical instability strip. It is important to note that the {\it observed} boundaries of the classical instability strip are purely empirical and are subsequently used to constrain theoretical models. Furthermore, the observed edges using ground-based observations of \dsct stars, which span $7400 \leq T_{\rm eff} \leq 8600$~K at $\log\,g \simeq 4.4$ \citep{Rod2001, Bowman_BOOK}, are narrower than the theoretical boundaries shown in Fig.~\ref{figure: Teff-logg}. Clearly, the theoretical red edge of the classical instability strip is more difficult to model compared to the blue edge because of the complex interaction of surface convection and the $\kappa$~mechanism in cool \dsct stars, with the convective envelope predicted to effectively damp pulsations in stars cooler than the red edge \citep{Pamyat1999a, Pamyat2000a, Dupret2004, Dupret2005, Houdek2015}. These results indicate that further theoretical work is needed in understanding mode excitation in \dsct stars, specifically those pulsating in higher radial orders ($n \geq 6$), and the large temperature range of observed \dsct stars. 

In summary, it is concluded that a statistically non-negligible fraction of \dsct stars are located cooler than the red edge and hotter than the blue edge of the classical instability strip calculated by \cite{Dupret2005}, as shown in Fig.~\ref{figure: Teff-logg}. A plausible explanation for the minority of \dsct stars located outside of the classical instability strip using the KIC and \citet{Huber2014} values are that they have incorrect $T_{\rm eff}$ and $\log\,g$ values, since these were determined using photometry, but outliers do exist with some stars having stellar parameters confirmed using spectroscopy. Therefore, these results demonstrate the need for a mass-dependent value of $\alpha_{\rm MLT}$ to reproduce all of the \dsct stars observed by the \Kepler Space Telescope. The \dsct stars that lie outside the classical instability strip will be interesting to study further, as they lie in a region where the excitation of low-radial order p~modes by the $\kappa$~mechanism is not expected to occur \citep{Pamyat1999a, Pamyat2000a, Dupret2004, Dupret2005}.


\section{Characterising the pulsation properties of \dsct stars observed by \Kepler}
\label{section: characterising}

In this section, correlations between the pulsation properties and stellar parameters of \dsct stars are investigated using the LC and SC ensembles. The pulsation properties of \dsct stars are diverse and, upon first inspection, little commonality is often found when comparing the amplitude spectra of individual \dsct stars in terms of the number of excited pulsation modes, their amplitudes and the observed frequency range. Since the number of significant pulsation modes in \dsct stars ranges from as small as a single mode\footnote{Examples of mono-periodic \dsct stars include KIC~5989856, KIC~8196006, KIC~8196381 and KIC~11509728, which were identified by \citet{Bowman2016a} as having only a single significant pulsation mode in the p mode frequency range.} to hundreds, it is difficult to characterise and compare the total pulsation energy in a systematic way for an ensemble of stars \citep{Bowman_PhD}. This is complicated by the fact that the majority of \dsct stars have variable pulsation mode amplitudes, with the total observable pulsational budget being unconserved on time scales of order a few years \citep{Bowman2014, Bowman2016a}. Also, in a classical pre-whitening procedure, it is important to exclude harmonics and combination frequencies when quoting the range of observed pulsation mode frequencies as these frequencies are caused by the non-linear character of the pulsation modes and do not represent independent pulsation mode frequencies \citep{Papics2012b, Kurtz2015b, Bowman_BOOK}.


	\subsection{Pulsation mode amplitudes}
	\label{subsection: Amax}

	Amongst the \dsct stars, a subgroup of high-amplitude \dsct (HADS) stars exists, which were originally defined by \citet{McNamara2000a} as \dsct stars with peak-to-peak light amplitudes exceeding 0.3~mag. HADS stars are also typically slowly-rotating ($v\sin\,i \lesssim 40$~km\,s$^{-1}$) and pulsate in the fundamental and/or first-overtone radial modes. These stars are known to be rare, making up less than one~per~cent of pulsating stars within the classical instability strip \citep{Lee2008}, but the physical cause of their differences to typical \dsct stars remains unestablished, if any exists \citep{Balona2016b}.
	
	\begin{figure}
	\centering
	\includegraphics[width=\columnwidth]{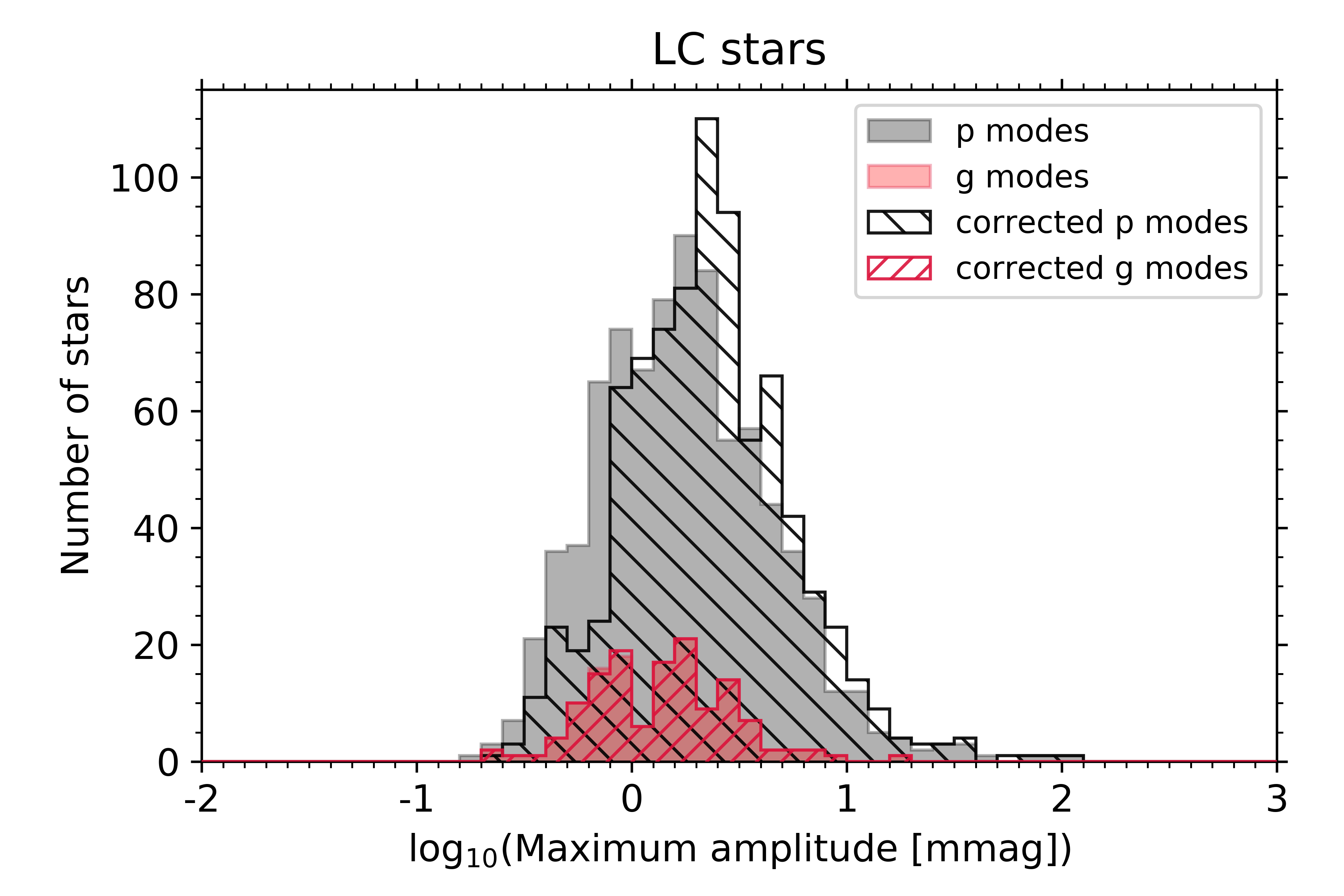}
	\includegraphics[width=\columnwidth]{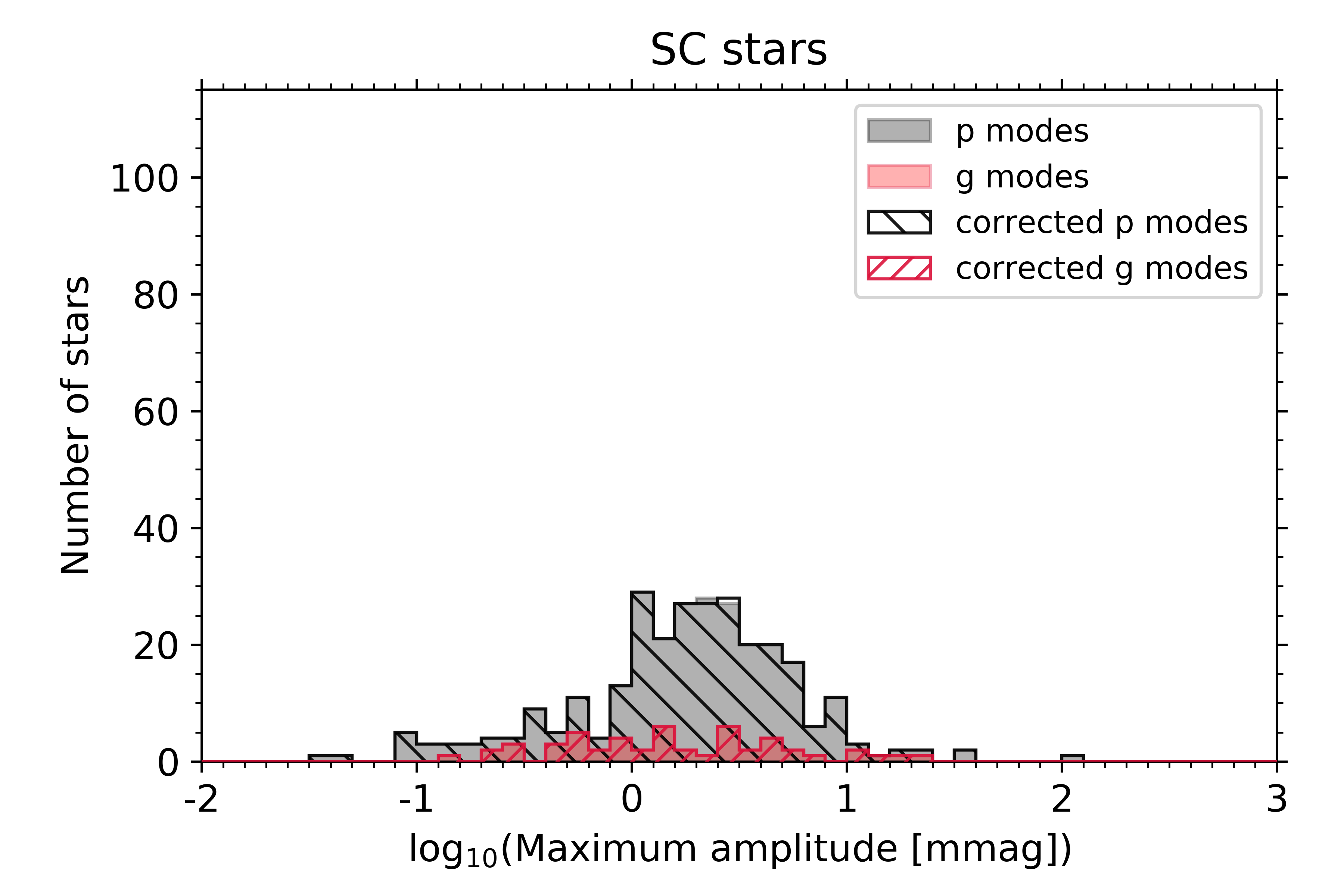}
	\caption{The distributions of the highest amplitude (i.e., dominant) pulsation mode for the ensemble of {963} LC and {334} SC \dsct stars are shown in the top and bottom panels, respectively. The black and red regions represent the stars for which the dominant pulsation mode is in the p-mode ($\nu \geq 4$~d$^{-1}$) or g-mode ($\nu < 4$~d$^{-1}$) frequency regimes, respectively. The filled regions represent the extracted pulsation mode amplitudes, with the hatched regions of the same colour representing each distribution after being corrected for the amplitude visibility function given in Eqn.~(\ref{equation: amp visibility}).}
	\label{figure: Amax}
	\end{figure}

	To demonstrate the rarity of HADS stars in the \Kepler mission data and examine the typical pulsation mode amplitudes in \dsct stars, the (logarithmic) distributions of the amplitude of the highest amplitude pulsation mode for the LC and SC ensembles are shown in the {top} and {bottom} panels of Fig.~\ref{figure: Amax}, respectively. In each panel of Fig.~\ref{figure: Amax}, the amplitudes of the dominant pulsation modes have been approximately separated into p~modes ($\nu \geq 4$~d$^{-1}$) and g~modes ($\nu < 4$~d$^{-1}$), and are showed as the filled grey and red areas respectively. It is possible for g-mode frequencies and combination frequencies of g~modes to exceed 4~d$^{-1}$, but typically they have lower amplitudes and do not represent the dominant pulsation mode in a star. Each amplitude is corrected for the amplitude visibility function given in Eqn.~\ref{equation: amp visibility}, with the corrected distributions for p and g~modes shown as the black- and red-hatched regions, respectively. As expected, the amplitude visibility function has a negligible effect on the SC ensemble. By separating \dsct stars into if their dominant pulsation mode corresponds to a g- or p-mode, it is demonstrated that a non-negligible fraction, more than 15~per~cent, of \dsct stars have their highest-amplitude pulsation mode in the g-mode frequency range. We also investigated using a value of 7~d$^{-1}$ to separate g-mode and p-mode dominated \dsct stars, which yielded similar results with a difference of order a few per~cent.
	
	 From inspection of Fig.~\ref{figure: Amax}, the amplitude of the dominant pulsation mode in a \dsct star typically lies in between {0.5} and {10} mmag. In each panel, a tail extending from moderate ($A \simeq 10$~mmag) to high pulsation mode amplitudes ($A \simeq 100$~mmag) can be seen, which is caused by a small number of HADS stars in each ensemble. Since the SC ensemble contains all \dsct stars irrespective of the length of data available, coupled with insignificant amplitude suppression described by the amplitude visibility function in Eqn.~(\ref{equation: amp visibility}), this demonstrates the intrinsic scarcity of HADS stars in the \Kepler data set \citep{Lee2008, Balona2016b, Bowman_BOOK}.

	It is known that the majority of \dsct stars exhibit amplitude modulation over the 4-yr \Kepler data set \citep{Bowman2016a, Bowman_PhD}, so it could be argued that a different {\it dominant} pulsation mode for each \dsct star could be extracted when using different subsets of \Kepler data. However, the amplitude distribution of \dsct stars would largely remain similar. This can be understood by comparing the top and bottom panels of Fig.~\ref{figure: Amax}, calculated using LC and SC \Kepler data, respectively. The SC observations of \dsct stars are randomly distributed throughout the 4-yr \Kepler mission, but the distributions in Fig.~\ref{figure: Amax} have similar median and FWHM values. This demonstrates that the pulsation properties of an ensemble of \dsct stars are less time-dependent than the pulsation properties of an individual \dsct star.


	\subsection{Pulsation and Effective Temperature}
	\label{subsection: freq-Teff}
	
	A semi-empirical relationship exists between the effective temperature and the observed pulsation mode frequencies in a \dsct star, which is caused by the depth of the He~{\sc ii} driving region in the stellar envelope being a function of $T_{\rm eff}$ \citep{C-D2000, ASTERO_BOOK}. A ZAMS \dsct star near the blue edge of the classical instability strip is expected to pulsate in higher radial overtone modes leading to higher observed pulsation mode frequencies compared to a ZAMS \dsct star near the red edge of the classical instability strip \citep{Pamyat1999a, Pamyat2000a, Dupret2004, Dupret2005}. This relationship between effective temperature and pulsation mode frequencies in \dsct stars has been previously been demonstrated using ground-based observations (see e.g., \citealt{Breger1975, Breger2000b, Rod2001}). Observationally, the blue and red edges are known for ground-based data \citep{Rod2001}, but the situation becomes less clear when the criterion of $S/N \geq 4$ is applied to a data set with a noise level of order a few $\mu$mag such as \Kepler mission data, as demonstrated in section~\ref{section: Instability Strip}.
	
	\begin{figure*}
	\centering
	\includegraphics[width=\columnwidth]{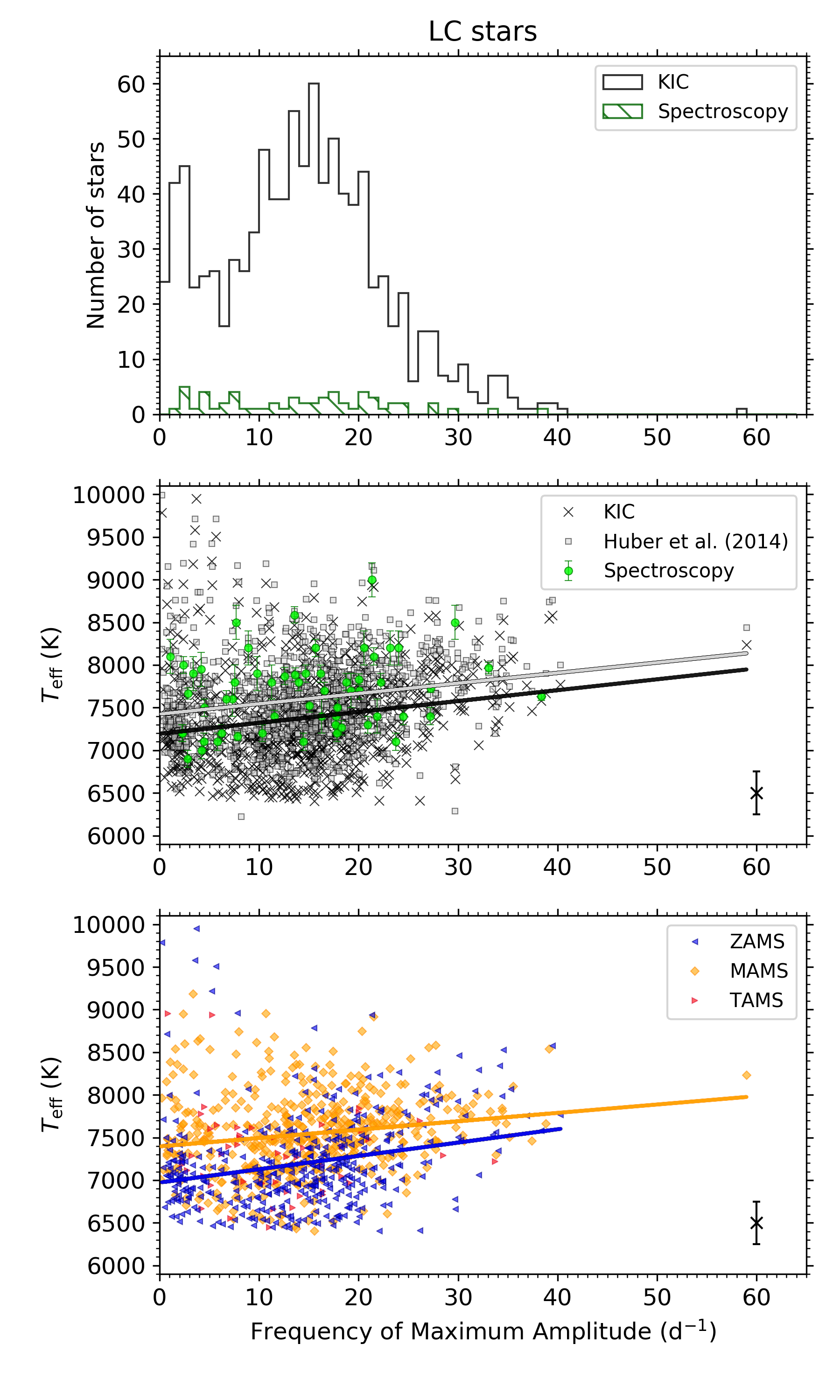}
	\includegraphics[width=\columnwidth]{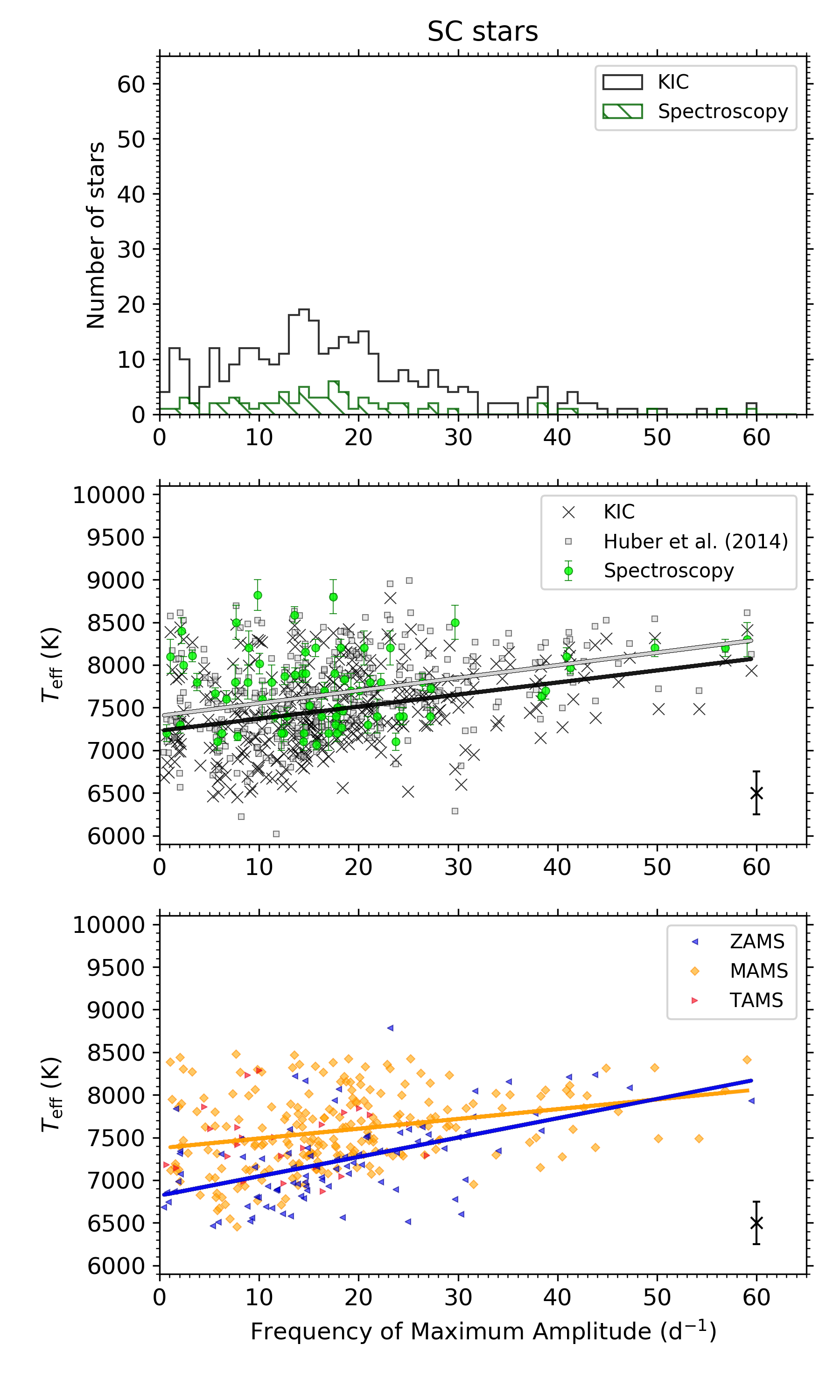}
	\caption{The top row shows the distribution of the frequency of maximum amplitude shown as the black region, with the subgroup of stars studied by \citet{Tkachenko2012a, Tkachenko2013a} and \citet{Niemczura2015, Niemczura2017a} plotted as the green-hatched area. The middle row shows the frequency of maximum amplitude versus the effective temperature, using the KIC, \citet{Huber2014} and spectroscopic values from \citet{Tkachenko2012a, Tkachenko2013a} and \citet{Niemczura2015, Niemczura2017a} as black crosses, grey squares and green circles, respectively, with statistically-significant linear regressions shown as solid lines. The bottom shows the same relationship as the middle row, but for only the KIC stars that have grouped based on their $\log\,g$ value into bins of $\log\,g \geq 4.0$ (ZAMS), $3.5 \leq \log\,g < 4.0$ (MAMS) and $\log\,g < 3.5$ (TAMS). For all rows, the left and right hand plots correspond to the LC and SC ensembles, respectively, which have been plotted using the same ordinate and abscissa scales for comparison.}
	\label{figure: freq}
	\end{figure*}
	
	The frequency of the dominant pulsation mode in each \dsct star in the LC and SC ensembles are shown in the left and right columns of Fig.~\ref{figure: freq}, respectively. In the top row, the distributions of the dominant pulsation mode frequency are shown by the black region, with the subgroup of stars studied by \citet{Tkachenko2012a, Tkachenko2013a} and \citet{Niemczura2015, Niemczura2017a} shown as the green-hatched region for comparison. Although the LC ensemble suffers from a bias caused by the amplitude visibility function, a dearth of \dsct stars with pulsation mode frequencies above 60~d$^{-1}$ exists in both ensembles, such that $0 < \nu \leq 60$~d$^{-1}$ can be considered the typical frequency range of pulsation modes of \dsct stars in the \Kepler data set. These results are not necessarily representative of {\it all} \dsct stars since the TAMS is better sampled compared to the ZAMS in the \Kepler data set, with more-evolved \dsct stars expected to have lower pulsation mode frequencies. The frequency of the dominant pulsation mode does not exceed 60~d$^{-1}$ in any \dsct star in either ensemble, which is in agreement with the expectations from theoretical models of main sequence \dsct stars near the blue edge of the classical instability strip \citep{Pamyat1999a, Pamyat2000a}.
	
	The frequency of the highest amplitude pulsation mode is plotted against $T_{\rm eff}$ for the ensemble of {963} LC and {334} SC \dsct stars in the middle row of Fig.~\ref{figure: freq}, in which the KIC parameters, the revised \citet{Huber2014} parameters and the subgroup of stars studied by \citet{Tkachenko2012a, Tkachenko2013a} and \citet{Niemczura2015, Niemczura2017a} are shown as black crosses, grey squares and green circles, respectively. Statistically significant linear regressions are also shown as solid lines, with the gradient, $m$, intercept, $c$, coefficient of correlation, $R$, and $p$-value obtained from a $t$-test for each data set given in Table~\ref{table: fit stats}. Linear regressions of the frequency of maximum amplitude and effective temperature using either the KIC or \citet{Huber2014} values yield statistically-significant positive correlations with $R \simeq 0.2$ and $R \simeq 0.3$ for the LC and SC ensembles, respectively. The amplitude suppression of high frequency pulsation modes in LC \Kepler data explains the weaker correlation between pulsation and effective temperature found in the LC ensemble compared to the SC ensemble, since few \dsct stars with pulsation mode frequencies above $40$~d$^{-1}$ are included. On the other hand, statistically-significant correlations between pulsation and effective temperature for stars in the spectroscopic subgroup were not found. From the relatively large $p$-values calculated from the linear regression in the spectroscopic subgroup, we cannot claim that the positive correlation found between effective temperature and frequency of the highest-amplitude pulsation mode is statistically significant. With few \dsct stars available with high-resolution spectroscopy, more are needed to investigate this relationship further.
	
	\begin{table*}
	\centering
	\caption{Statistics for linear regressions of the frequency of maximum amplitude and effective temperature for the \dsct stars in this paper, using KIC values \citep{Brown2011}, the revised KIC values from \citet{Huber2014} and spectroscopic values from \citet{Tkachenko2012a, Tkachenko2013a} and \citet{Niemczura2015, Niemczura2017a}. The column headers include the number of stars, $N_{\rm stars}$, the gradient, $m$, the ordinate-axis intercept of the linear fit, $c$, the coefficient of correlation, $R$ and the corresponding $p$-value (obtained from a $t$-test). Separate regressions for the $\log\,g$ subgroups are given for KIC and \citet{Huber2014} values.}
	\begin{tabular}{l r r r r r c r r r r r}
	\hline
	& \multicolumn{5}{c}{Long Cadence (LC) ensemble} && \multicolumn{5}{c}{Short Cadence (SC) ensemble} \\
	& \multicolumn{1}{c}{$N_{\rm stars}$} & \multicolumn{1}{c}{$m$} & \multicolumn{1}{c}{$c$} & \multicolumn{1}{c}{$R$} & \multicolumn{1}{c}{$p$-value} &  \multicolumn{1}{c}{$ $} & \multicolumn{1}{c}{$N_{\rm stars}$} & \multicolumn{1}{c}{$m$} & \multicolumn{1}{c}{$c$} & \multicolumn{1}{c}{$R$} & \multicolumn{1}{c}{$p$-value} \\
	& \multicolumn{1}{c}{} & \multicolumn{1}{c}{(K / d$^{-1}$)} & \multicolumn{1}{c}{(K)} & \multicolumn{1}{c}{} & \multicolumn{1}{c}{} & \multicolumn{1}{c}{} & \multicolumn{1}{c}{} & \multicolumn{1}{c}{(K / d$^{-1}$)} & \multicolumn{1}{c}{(K)} & \multicolumn{1}{c}{} & \multicolumn{1}{c}{} \\
	\hline
	\multicolumn{1}{c}{KIC} \\
	
	All						&	{963}	&	{$12.8 \pm 2.0$}	&	{$7195 \pm 17$}	&	{0.20}	&	{$< 0.0001$}	&&	{334}	&	{$14.1 \pm 2.2$}	&	{$7229 \pm 21$}	&	{0.33}	& 	{$< 0.0001$}	\\
	
	ZAMS ($\log\,g \geq 4.0$)		&	{421}	&	{$15.6 \pm 3.0$}	&	{$6972 \pm 16$}	&	{0.24}	&	{$< 0.0001$}	&&	{89}		&	{$22.7 \pm 4.0$}	&	{$6819 \pm 21$}	&	{0.52}	& 	{$< 0.0001$}	\\
	
	MAMS ($3.5 \leq \log\,g < 4.0$)	&	{494}	&	{$9.8 \pm 2.5$}		&	{$7396 \pm 17$}	&	{0.17}	&	{$0.0001$}	&&	{221}	&	{$11.4 \pm 2.5$}	&	{$7375 \pm 22$}	&	{0.29}	& 	{$< 0.0001$}	\\
	
	TAMS ($\log\,g < 3.5$)		&	{48}		&	{$-7.7 \pm 9.9$}	&	{$7404 \pm 14$}	&	{-0.11}	&	{$0.4438$}	&&	{24}		&	{$2.3 \pm 11.5$}	&	{$7442 \pm 14$}	&	{0.04}	& 	{$0.8460$}	\\
	
	\hline
	\multicolumn{1}{c}{Revised KIC} \\
	
	All						&	{963}	&	{$12.1 \pm 2.0$}	&	{$7424 \pm 17$}	&	{0.19}	&	{$< 0.0001$}	&&	{329}	&	{$14.8 \pm 2.2$}	&	{$7404 \pm 21$}	&	{0.34}	& 	{$< 0.0001$}	\\
	
	ZAMS ($\log\,g \geq 4.0$)		&	{423}	&	{$15.5 \pm 2.9$}	&	{$7215 \pm 17$}	&	{0.25}	&	{$< 0.0001$}	&&	{85}		&	{$23.0 \pm 4.2$}	&	{$7027 \pm 21$}	&	{0.52}	& 	{$< 0.0001$}	\\
	
	MAMS ($3.5 \leq \log\,g < 4.0$)	&	{506}	&	{$9.1 \pm 2.6$}		&	{$7606 \pm 17$}	&	{0.16}	&	{$0.0004$}	&&	{224}	&	{$13.1 \pm 2.5$}	&	{$7512 \pm 21$}	&	{0.33}	& 	{$< 0.0001$}	\\
	
	TAMS ($\log\,g < 3.5$)		&	{34}		&	{$-0.5 \pm 12.9$}	&	{$7525 \pm 15$}	&	{-0.01}	&	{$0.9710$}	&&	{20}		&	{$4.6 \pm 12.2$}	&	{$7662 \pm 17$}	&	{0.09}	&	{$0.7114$}	\\	
	
	\hline
	\multicolumn{1}{c}{Spectroscopy} \\
	
	All 						&	{58}		&	{$11.4 \pm 6.7$}	&	{$7515 \pm 17$}	&	{0.22}	&	{$0.0947$}	&&	{70}		&	{$6.2 \pm 4.4$}		&	{$7625 \pm 21$}	&	{0.17}	& 	{$0.1584$}	\\
	
	\hline
	\end{tabular}
	\label{table: fit stats}
	\end{table*}
	
	However, it is important to distinguish that the expectation of hotter \dsct stars having higher pulsation mode frequencies is predicted for ZAMS stars, with the relationship breaking down for TAMS and post-main sequence \dsct stars. As for any star, the frequencies of its pulsation modes decrease, albeit slowly, during the main sequence because of the increase in stellar radius. This creates a degeneracy when studying an ensemble of stars, with pulsation mode frequencies expected to be correlated with $T_{\rm eff}$ but also correlated with $\log\,g$ (i.e., inversely with age on the main sequence). In practice, the sensible approach is to test the relationship between the frequency of the dominant pulsation mode and effective temperature for different groups of stars based on their $\log\,g$ value. This is shown in the bottom row of Fig.~\ref{figure: freq} for stars in the LC and SC ensembles that have grouped into $\log\,g \geq 4.0$ (ZAMS), $3.5 \leq \log\,g < 4.0$ (Mid-Age Main Sequence; MAMS) and $\log\,g < 3.5$ (TAMS) using their KIC parameters. 
	
	Linear regressions for each of these $\log\,g$ subgroups for the KIC and \citet{Huber2014} values are also included in Table~\ref{table: fit stats}, which demonstrate the need to differentiate ZAMS and TAMS stars when studying the pulsational properties of \dsct stars. For example, as shown in Fig.~\ref{figure: freq} and given in Table~\ref{table: fit stats}, ZAMS stars in the LC and SC ensembles show a statistically-significant positive correlation between effective temperature and frequency of the highest-amplitude pulsation, with values of $R \simeq 0.3$ and $R \simeq 0.5$, respectively. As discussed previously, the lack of high-frequency pulsators in the LC ensemble explains the difference in these two values. Conversely, no statistically-significant correlation was found for TAMS \dsct stars in either the LC or SC ensembles.
	
	Linear regressions for subgroups based on $\log\,g$ for the \dsct stars with spectroscopic parameters from \citet{Tkachenko2012a, Tkachenko2013a} and \citet{Niemczura2015, Niemczura2017a} do not provide reliable results because most of these stars have surface gravities between $3.5 \leq \log\,g \leq 4.0$, and because no significant correlation was found for the subgroup as a whole. Clearly, the ZAMS \dsct stars, shown in blue in the bottom row of Fig.~\ref{figure: freq}, have the strongest statistical correlation between $T_{\rm eff}$ and frequency of maximum amplitude with $R \simeq 0.5$, which is in agreement with theoretical models and previous observations \citep{Pamyat1999a, Pamyat2000a, Breger2000b, Rod2001, Dupret2004, Dupret2005, Houdek2015}.


	\subsection{Low-frequency variability in \dsct stars}
	\label{subsection: low-freq}
	
	The incidence of g~modes in stars that exceed $T_{\rm eff} \gtrsim 8000$~K, especially those with accurate stellar parameters obtained from high-resolution spectroscopy, are interesting cases. These stars are significantly hotter than typical \gdor stars observed by {\it Kepler}, with effective temperatures of \gdor stars determined using high-resolution spectroscopy typically between $6900 \leq T_{\rm eff} \leq 7400$~K \citep{Tkachenko2013a, VanReeth2015a}. The excitation of g~modes in \dsct stars that are located hotter than the blue edge of the \gdor instability region warrants further study, with such stars found in both LC and SC ensembles. These hot, low-frequency pulsators can be clearly seen in each panel of Fig.~\ref{figure: freq}, in which a non-negligible fraction of stars (originally chosen because they contain only p modes, or p and g~modes) have their dominant pulsation mode in the g~mode frequency regime. 
	
	A distinct bi-modality is present in the top-left panel of Fig.~\ref{figure: freq} with a minimum in the distribution at a frequency of approximately 7~d$^{-1}$. This minimum in the distribution corresponds to the approximate upper limit of g-mode pulsation frequencies in fast-rotating stars \citep{Bouabid2013, VanReeth2016a, Aerts2017b} and an approximate lower limit of p-mode pulsation frequencies, although overlap of g and p-modes and their combination frequencies is expected between $4 \leq \nu \leq 7$~d$^{-1}$. Therefore, the stars with the dominant pulsation mode in the g-mode frequency-regime shown in Fig.~\ref{figure: freq} represent hybrid stars in which the dominant pulsation mode corresponds to a g~mode. However, it remains unclear how or why there should be a significant difference, if any, in the respective amplitudes of the p and g~modes in a \dsct star, with all stars in both ensembles selected to contain at least p~modes.
	
	The \Kepler mission data have proven extremely useful for studying the interior physics of intermediate-mass stars, specifically concerning measurements of radial rotation and angular momentum transport \citep{Aerts2017b}. The first examples of hybrid main sequence A and F stars with measured interior rotation were by \citet{Kurtz2014} and \citet{Saio2015b}, and exemplify the power of asteroseismology when applied to hybrid stars such that model-independent measurements of rotation inside stars can be made. To date, a few dozen intermediate-mass main sequence stars have been found to be almost rigidly rotating or have weak differential rotation \citep{Degroote2010a, Kurtz2014, Papics2014, Saio2015b, VanReeth2015b, Triana2015, Murphy2016a, Ouazzani2017a, Papics2017a, Zwintz2017a, VanReeth2018a**}.
	
	The full asteroseismic potential of studying the hybrid stars using the high-quality \Kepler data has yet to be exploited. For example, the observed power excess at low frequencies in \gdor stars has recently been interpreted as Rossby modes, which further inform our understanding of rotation \citep{VanReeth2016a, Saio2018a}. Furthermore, expanding the systematic search for period spacing patterns beyond the \gdor stars to the hybrid stars has the prospect of extending the studies of rotation and angular momentum transport to higher masses and fill the gap between the published cases of B and F~stars in the literature. The hybrid stars in the LC ensemble have the necessary data quality to extract, identify and model g and p~modes which provide valuable physical constraints of the near-core and near-surface regions within a star, respectively, with more observational studies of intermediate- and high-mass stars needed to address the large shortcomings in the theory of angular momentum transport when comparing observations of main sequence and evolved stars \citep{Tayar2013, Cantiello2014, Eggenberger2017a, Aerts2017b}.
	
	The study of angular momentum transport in stars has revealed that Internal Gravity Waves (IGWs) stochastically-driven in stars with convective cores can explain the observed near-rigid rotation profiles in intermediate-mass stars \citep{Rogers2013b, Rogers2015}. The detection of IGWs can be inferred by the power-excess at low frequencies in a star's amplitude spectrum. The few observational detections of IGWs in the literature have been for massive stars of spectral type O, and were made by matching the observed morphology of the low-frequency power excess with that predicted by state-of-the-art simulations of IGWs \citep{Rogers2013b, Rogers2015, Aerts2015c, Aerts2017a, Aerts2018a*, Simon-Diaz2018a*}. Any star with a convective core is predicted to excite IGWs, thus the ensemble of \dsct stars presented in this work may contain observational signatures of IGWs (e.g., \citealt{Tkachenko2014a}), and warrant further study. 
	
	The \Kepler data set will remain unchallenged for the foreseeable future for studying \dsct stars because of its total length, duty cycle and photometric precision. Furthermore, it represents a homogeneous data set for detections of IGWs and the study of mixing and angular momentum transport in intermediate-mass A and F stars.


	\subsection{Pulsation within the Classical Instability Strip}
	\label{subsection: HR}
	
\begin{figure*}
	\centering
	\includegraphics[width=\columnwidth]{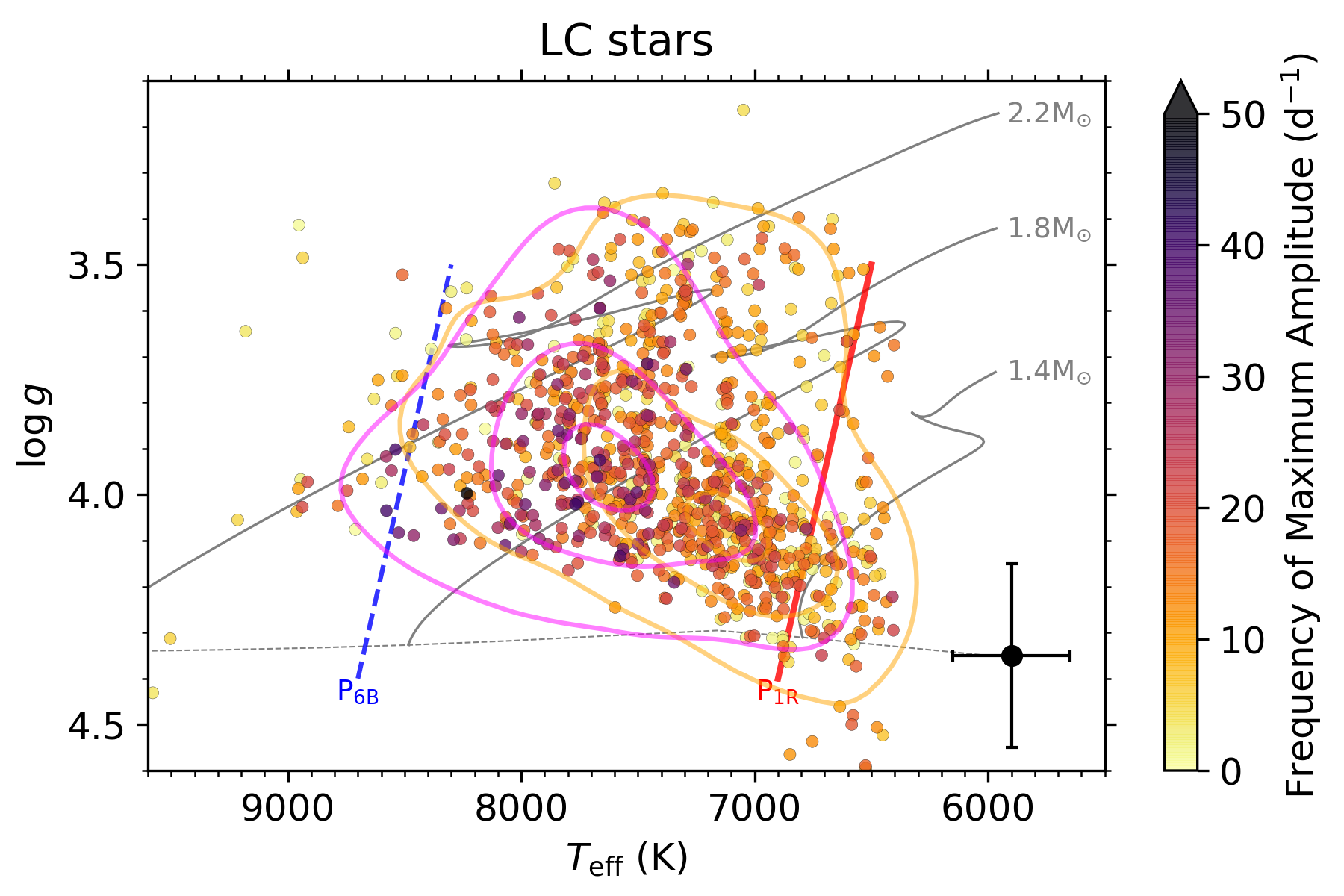}
	\includegraphics[width=\columnwidth]{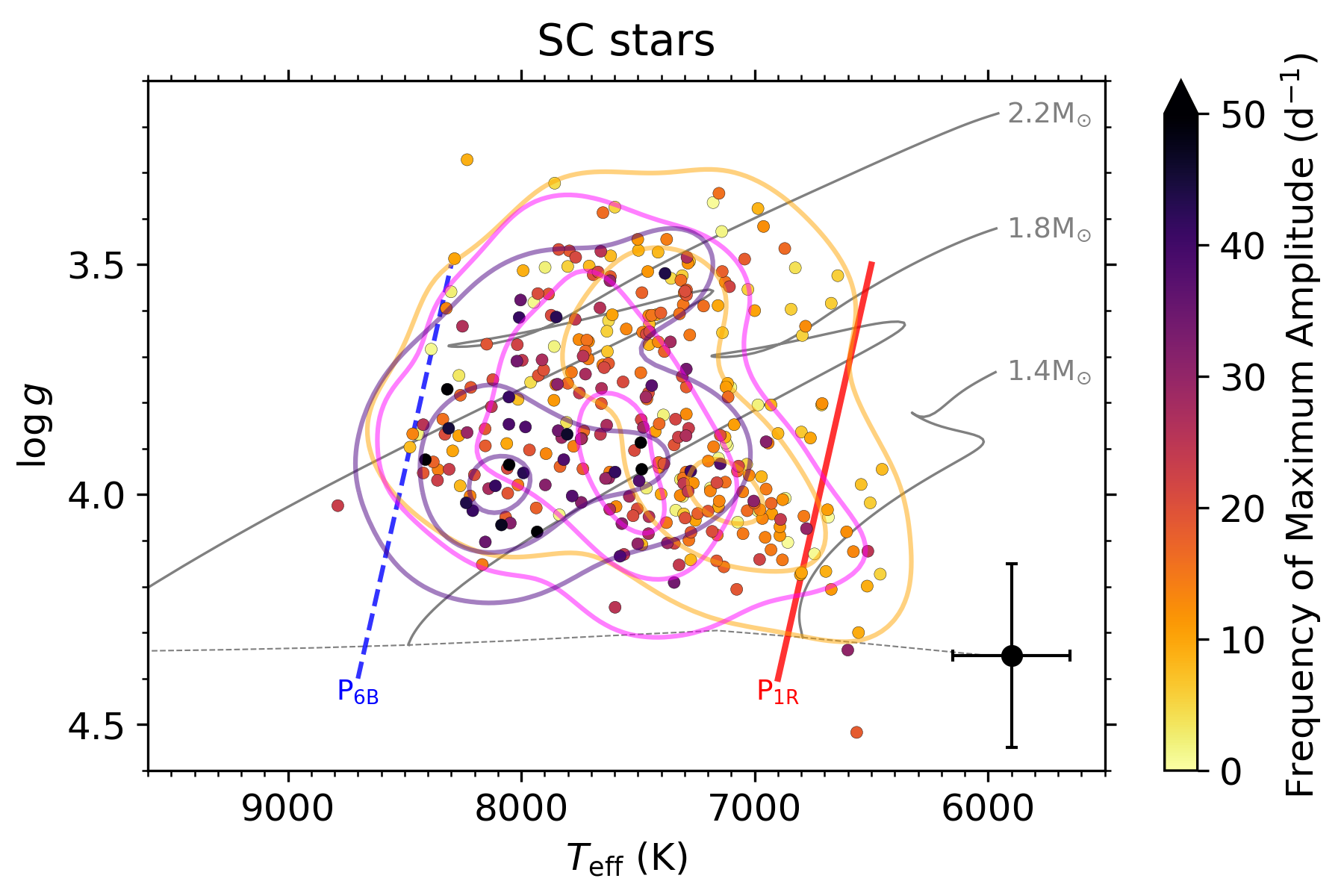}
	\includegraphics[width=\columnwidth]{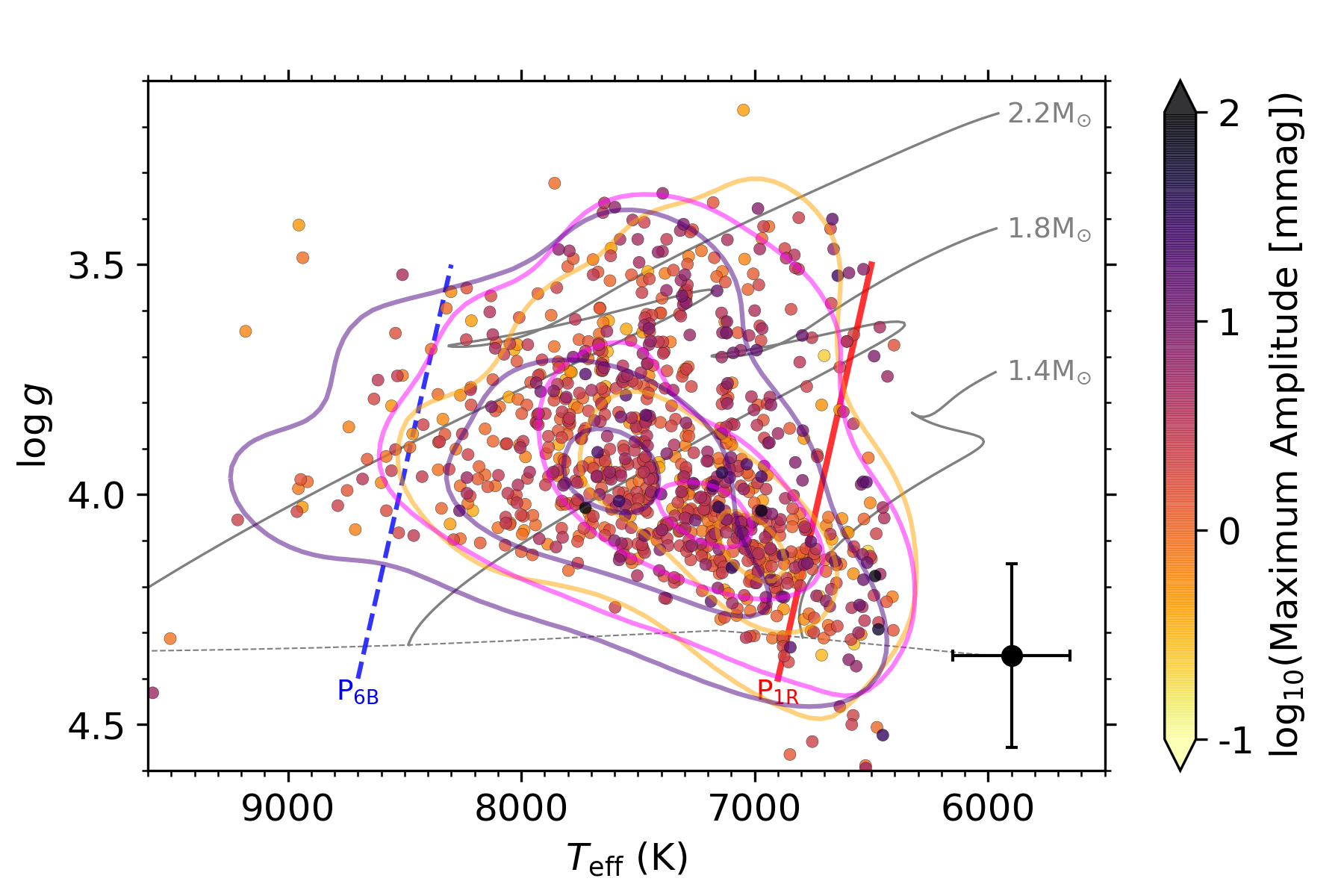}
	\includegraphics[width=\columnwidth]{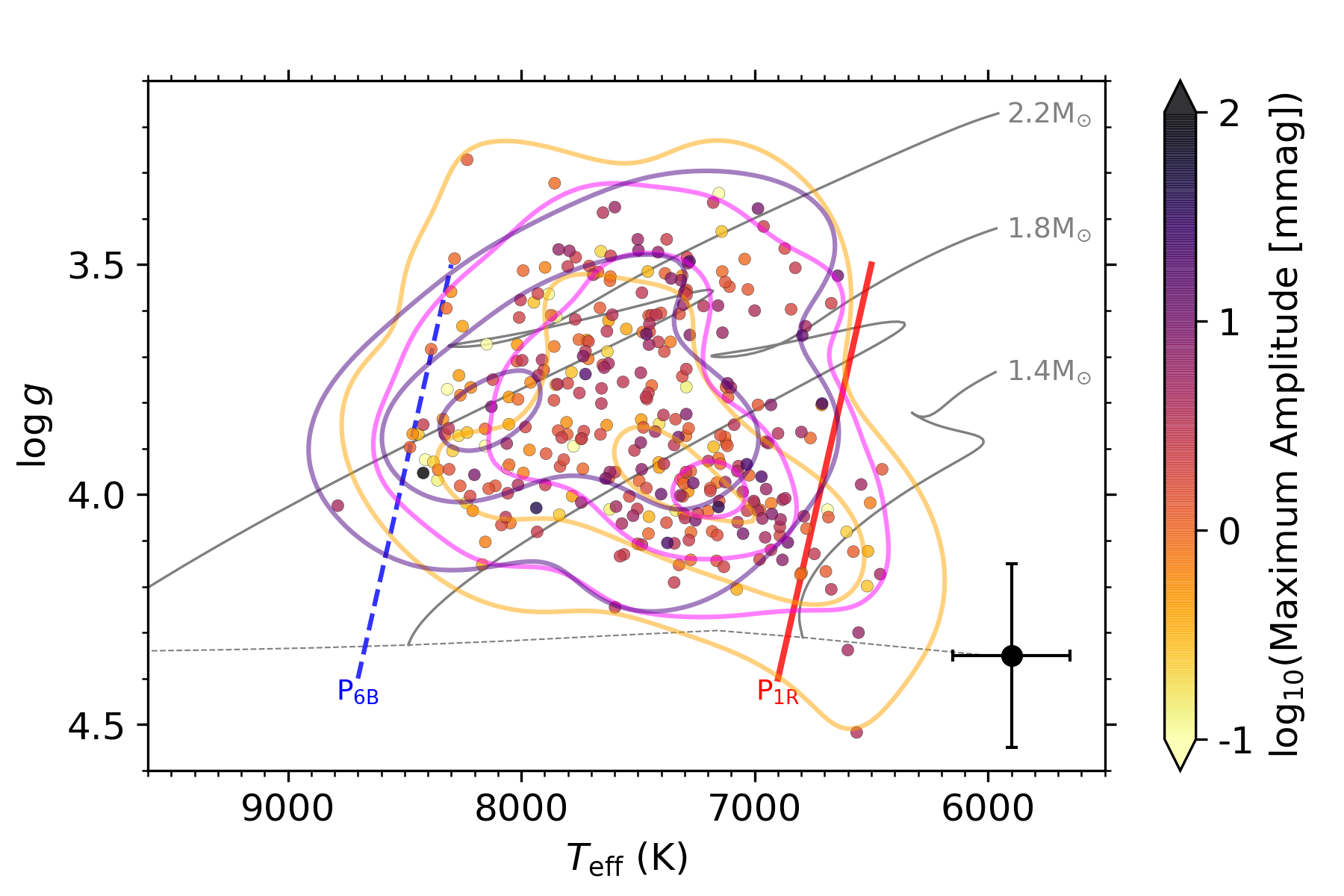}
	\caption{The left and right columns correspond to the LC and SC ensembles of \dsct stars, respectively. The top row shows the location of each \dsct in a  $T_{\rm eff}-\log\,g$ diagram using the KIC parameters, with each star's location colour-coded by the frequency of the highest-amplitude pulsation mode. Similarly, the bottom row is colour-coded by the amplitude of the highest-amplitude pulsation mode using a logarithmic colour-bar scale. The same stellar evolutionary tracks, ZAMS line, typical KIC error bar and theoretical edges of the classical instability strip as in Fig.~\ref{figure: Teff-logg} are also shown. In each panel, density contours are plotted for the stars contained within three parts of the shown colour bar scale.}
	\label{figure: Teff-logg colour}
	\end{figure*}

	The relationships between pulsation mode frequencies and effective temperature and evolutionary stage (i.e., $\log\,g$) are degenerate, with pulsation mode frequencies predicted and observed to be correlated with both effective temperature and surface gravity. Therefore, using a $T_{\rm eff}-\log\,g$ diagram is a sensible way to investigate these correlations further. The LC and SC ensembles of \dsct stars are once again plotted in $T_{\rm eff}-\log\,g$ diagrams in the left and right columns of Fig.~\ref{figure: Teff-logg colour}, respectively, using the KIC parameters for each star. However, unlike the previous diagrams (shown in Fig.~\ref{figure: Teff-logg}), each star in Fig.~\ref{figure: Teff-logg colour} is plotted as a filled circle that has been colour-coded by the frequency of its highest-amplitude pulsation mode in the top row, and colour-coded by the amplitude of its highest-amplitude pulsation mode in the bottom row. 
		
	In each panel in Fig.~\ref{figure: Teff-logg colour}, contour lines corresponding to the $10^{\rm th}$, $50^{\rm th}$ and $90^{\rm th}$ percentiles of the normalised probability density after applying a Gaussian kernel to the distribution of stars are shown for three subgroups of stars that fall into three parts of the colour bar scale. For example, the density maxima in the top-right panel of Fig.~\ref{figure: Teff-logg colour} correspond to approximate $T_{\rm eff}$ values of 7000~K, 7500~K and 8000~K for the $0 < \nu_{\rm max} < 20$~d$^{-1}$, $20 \leq \nu_{\rm max} < 40$~d$^{-1}$ and $\nu_{\rm max} \geq 40$~d$^{-1}$ subgroups, respectively. Note that since there are so few high-frequency pulsators in the LC ensemble, only two sets of contours are shown in the top-left panel of Fig.~\ref{figure: Teff-logg colour}. For the maximum amplitude distributions shown in the bottom row of Fig.~\ref{figure: Teff-logg colour}, the three sets of contours correspond to $\log_{10}(A_{\rm max}) \leq 0$, $ 0 < \log_{10}(A_{\rm max}) < 1$ and $\log_{10}(A_{\rm max}) \geq 1$. The choice of these subgroups is somewhat arbitrary, but they demonstrate the average trends in each panel and help to guide the eye compared to the distribution using individual stars.
		
	As expected and demonstrated using a subsample of ZAMS stars ($\log\,g$ > 4.0) in Fig.~\ref{figure: freq}, the \dsct stars with higher pulsation mode frequencies show higher effective temperatures and are found closer to the blue edge of the classical instability strip. The same relationship between effective temperature and pulsation mode frequencies can be seen in the top row of Fig.~\ref{figure: Teff-logg colour}, with high-frequency \dsct stars (shown in dark purple) being typically located nearer the blue edge of the classical instability strip. The situation is somewhat less clear when studying the relationship between the amplitude of the dominant pulsation mode and location in the $T_{\rm eff}-\log\,g$, which is shown in the bottom row of Fig.~\ref{figure: Teff-logg colour}. In the LC ensemble, the highest-amplitude pulsators (shown in dark purple) are more centrally located in the classical instability strip, whereas the high-amplitude pulsators in the SC ensemble are located closer to the blue edge, but this discrepancy is likely caused by so few high-amplitude stars in each ensemble. It has been previously demonstrated using ground-based observations that HADS stars are often found in the centre of the classical instability strip \citep{McNamara2000a, Breger2000b}, but with so few HADS stars in the \Kepler data set and the uncertainties in $T_{\rm eff}$ and $\log\,g$ values, it is difficult to make any meaningful conclusions.
		
	It should be noted that these results may not be completely representative of {\it all} \dsct stars, as only a single pulsation mode frequency was used to characterise each star. This in turn may explain the lack of significant correlation found between the frequency of maximum amplitude, $\nu_{\rm max}$, and effective temperature for all subgroups except ZAMS stars, as described by the results in Table~\ref{table: fit stats}. Using space-based observations such as those from {\it Kepler}, many \dsct stars have a large range of pulsation mode frequencies with small amplitudes. The density and range of the observed pulsation mode frequencies are not accounted for when extracting only a single pulsation mode --- this is explored in more detail in section~\ref{section: regularities}. Nonetheless, the high-frequency \dsct stars are typically located nearer the blue edge of the instability strip as shown in Fig.~\ref{figure: Teff-logg colour}, which is in agreement with theoretical predictions and previous observations of \dsct stars \citep{Breger1975, Pamyat1999a, Pamyat2000a, Breger2000b, Rod2001}. An interesting result from this analysis is that \dsct stars across the instability strip are able to pulsate in low and high pulsation mode frequencies, with no obvious explanation for which pulsation modes are excited in \dsct stars. Further observational and theoretical study of this is needed, as it will provide constrains of mode selection mechanism(s) at work in these stars.


	\subsection{Pulsation and rotation}
	\label{subsection: rotation}
	
	\begin{figure*}
	\centering
	\includegraphics[width=\columnwidth]{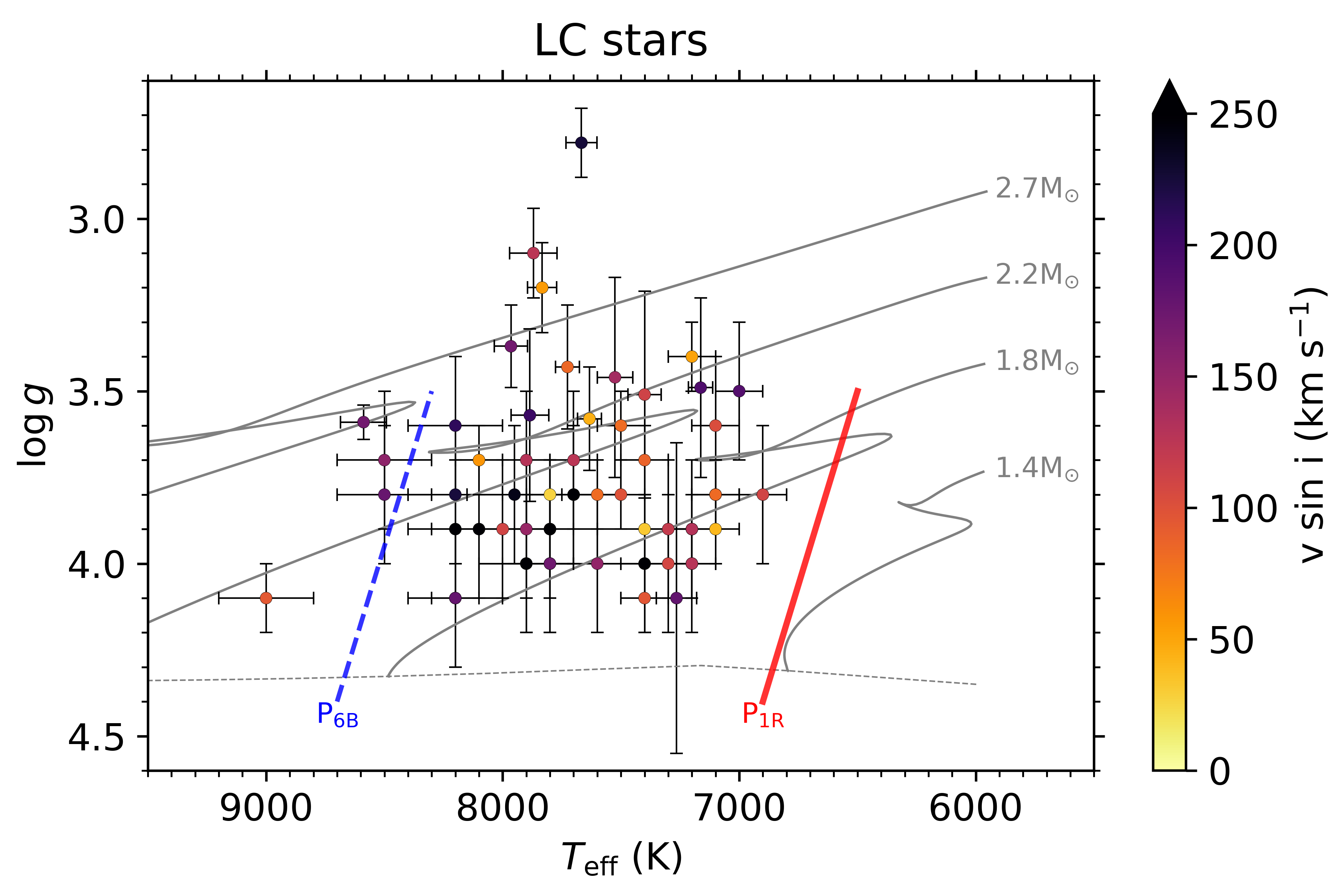}
	\includegraphics[width=\columnwidth]{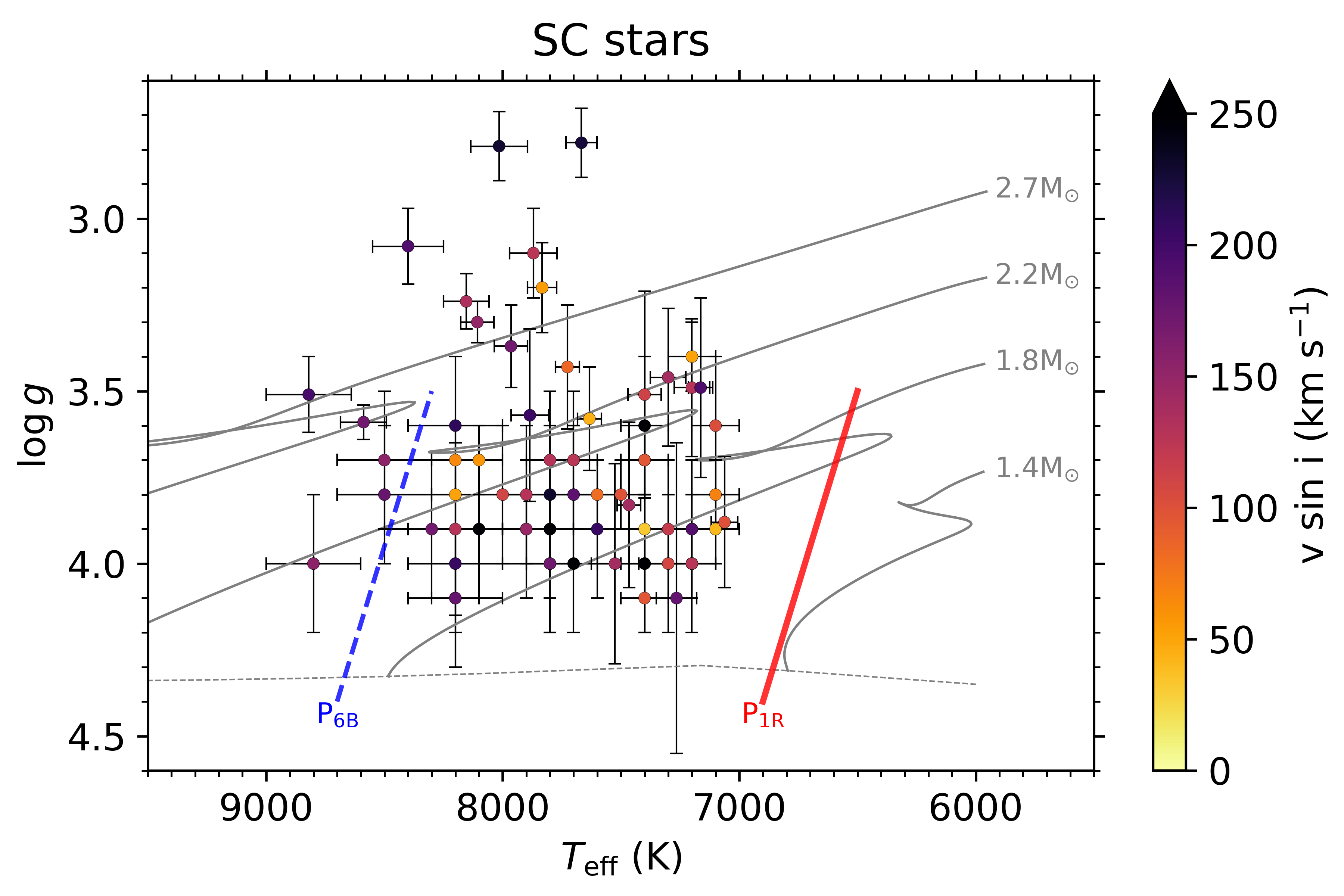}
	\includegraphics[width=\columnwidth]{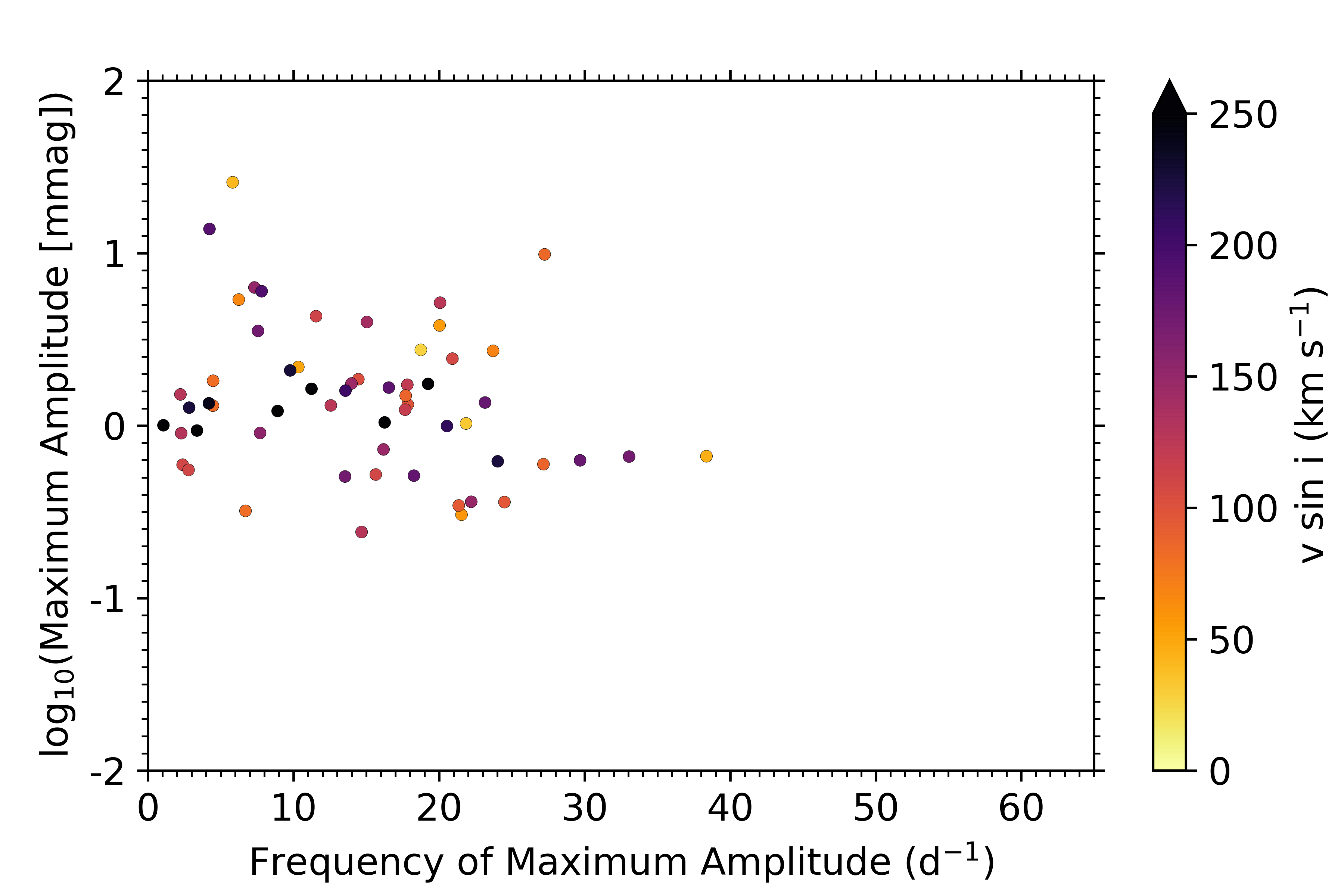}
	\includegraphics[width=\columnwidth]{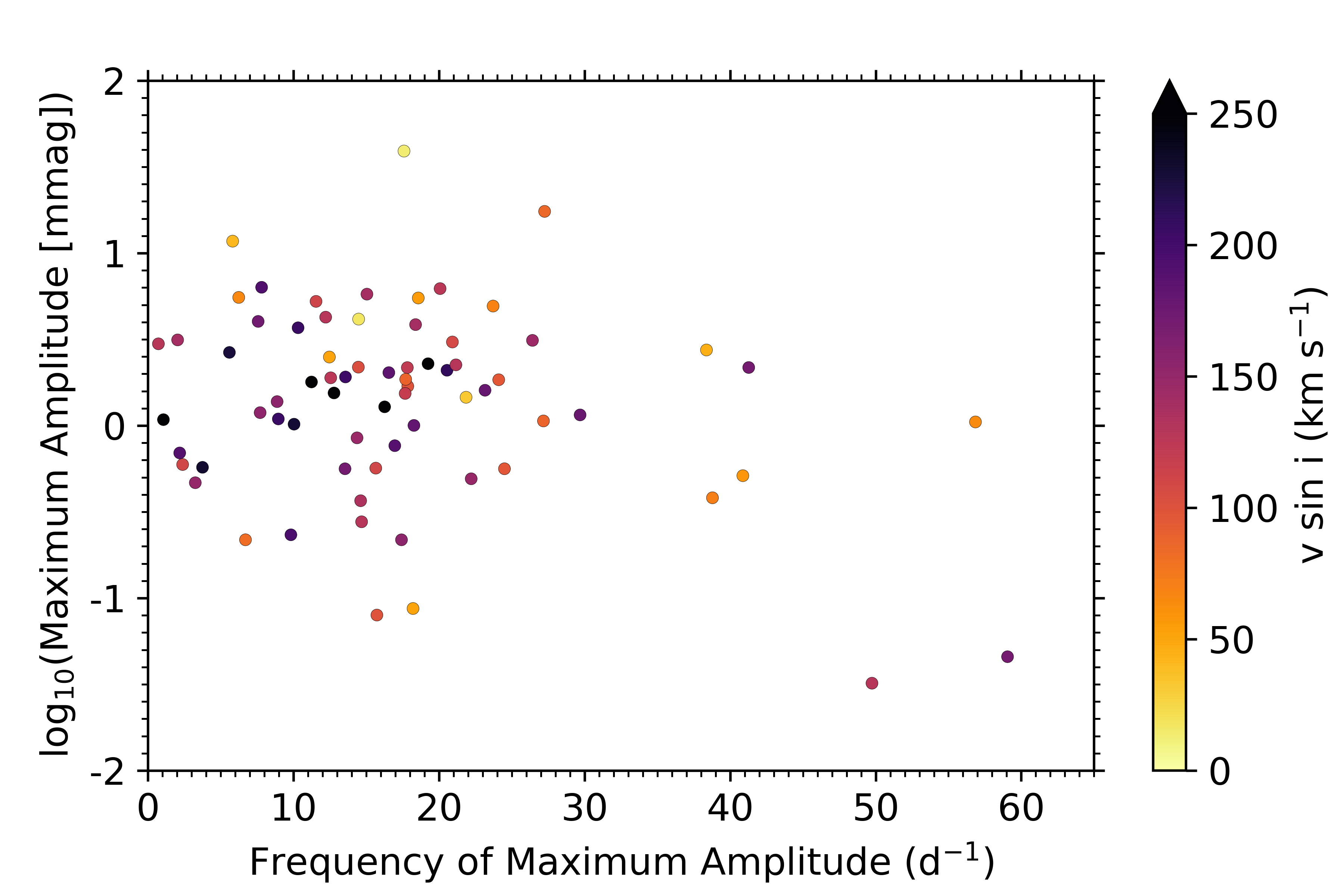}
	\caption{The left and right columns correspond to the LC and SC ensembles of \dsct stars, respectively, for which spectroscopic values of $T_{\rm eff}$, $\log\,g$ and $v\sin\,i$ are available. The top row shows the location of each \dsct in a  $T_{\rm eff}-\log\,g$ diagram using spectroscopic values from \citet{Tkachenko2012a, Tkachenko2013a} and \citet{Niemczura2015, Niemczura2017a}, with each star's location colour-coded by $v\sin\,i$. The same stellar evolutionary tracks, ZAMS line and theoretical edges of the classical instability strip as in Fig.~\ref{figure: Teff-logg} are also shown in the top row. The bottom row shows the relationship between the frequency and amplitude of the dominant pulsation mode colour-coded by $v\sin\,i$.}
	\label{figure: rotation}
	\end{figure*}
	
	The effects of rotation modify the pulsation mode frequencies of a star by lifting the degeneracy of non-radial modes into their $2\ell + 1$ components \citep{ASTERO_BOOK}, which are observed as multiplets in a star's amplitude spectrum. The splitting of these component frequencies is significantly asymmetric for moderate and fast rotating stars, with the Coriolis force acting against the direction of rotation and affects prograde and retrograde pulsation modes differently \citep{ASTERO_BOOK}. For a star with numerous non-radial pulsation modes, as is the case for numerous \dsct stars, the effects of rotation act to increase the observed range of frequencies in an amplitude spectrum (see e.g., \citealt{Breger2012b}). However, there is no expectation for the dominant pulsation mode frequency and/or amplitude of a \dsct star to be correlated with rotation.
		
	In Fig.~\ref{figure: rotation}, the subgroups of LC and SC \dsct stars that have stellar parameters from \citet{Tkachenko2012a, Tkachenko2013a} and \citet{Niemczura2015, Niemczura2017a} are shown in $T_{\rm eff} - \log\,g$ diagrams in the left and right panels, respectively. The top row shows the location of each star in a $T_{\rm eff} - \log\,g$ diagram with each star shown as a filled circle colour-coded by $v\sin\,i$ determined from spectroscopy. There is no clear correlation amongst effective temperature, surface gravity and rotation, with slow and fast rotators found across the classical instability strip. Of course, since the spectroscopic values of $v\,\sin\,i$ are projected surface rotational velocities, they represent lower limits of the true rotation of the stars shown in Fig.~\ref{figure: rotation}.
	
	In the bottom row of Fig.~\ref{figure: rotation}, the relationship between the frequency and amplitude of the dominant pulsation mode is shown, with each star colour-coded by $v\sin\,i$. This figure supports the findings of \citet{Breger2000a} that \dsct stars with high pulsation mode amplitudes are typically slow rotators ($v\sin\,i \lesssim 50$~km\,s$^{-1}$), but clearly not all slowly rotating \dsct stars have high pulsation mode amplitudes. However, the lack of any significant correlation in the panels of Fig.~\ref{figure: rotation} is not surprising with so few \dsct stars having been observed with high-resolution spectroscopy.


\section{Regularities in the amplitude spectra of \dsct star stars}
\label{section: regularities}

In this section, a similar methodology as that employed by \citet{Michel2017b}, who searched for regularities in the amplitude spectra of approximately 1800 \dsct stars observed by CoRoT, is applied to the SC ensemble of \dsct stars presented in this work. It is not informative to perform this analysis using the LC ensemble of \dsct stars because of the significantly longer integration times of LC \Kepler data suppressing high-frequency signals. Consequently, few \dsct stars exist in the LC ensemble that satisfy the required selection criteria of having high-frequency pulsation modes --- i.e., young \dsct stars. The SC \Kepler data were not obtained using as short a cadence as the CoRoT data, but it is still significantly higher than the pulsation mode frequencies in \dsct stars.
		
\begin{figure*}
\centering
\includegraphics[width=\textwidth]{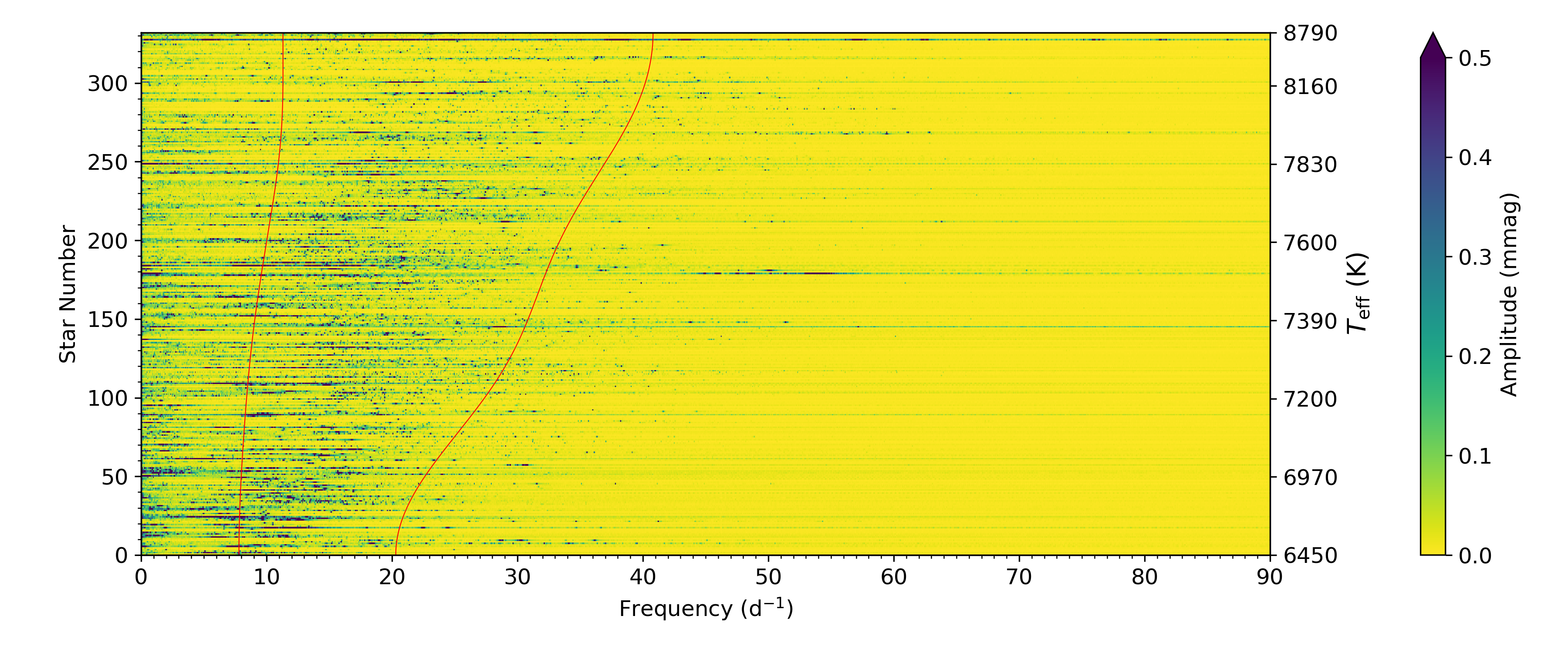}
\caption{The stacked amplitude spectra for the ensemble of SC \dsct stars in order of increasing $T_{\rm eff}$ using KIC values in an upwards direction, with the red lines indicating a moving average of the $\nu_{\rm low}$ and $\nu_{\rm high}$ values along the star number ordinate axis. Indicative effective temperature values are also included on the ordinate axis to demonstrate how $\nu_{\rm low}$ and $\nu_{\rm high}$ vary with $T_{\rm eff}$.}
\label{figure: FT contour 1}
\end{figure*}

Following \citet{Michel2017b}, an amplitude significance criterion was chosen as ten times the mean amplitude in each star's amplitude spectrum, with the frequency range of $\nu_{\rm high}$ and $\nu_{\rm low}$ calculated as the frequency of the highest- and lowest-amplitude peaks that satisfied the amplitude significance criteria for frequencies above $\nu \geq 4$~d$^{-1}$ ($\gtrsim 50~\mu$Hz). This minimum in frequency is approximately twice the value chosen by \citet{Michel2017b}, but we use 4~d$^{-1}$ in this study to reduce the chance that the extracted values of $\nu_{\rm low}$ represent g-mode pulsation frequencies. Each star's amplitude spectrum was interpolated onto a frequency resolution of 0.05~d$^{-1}$ using the maximum amplitude value of each bin in the original spectrum. This downgrades each amplitude spectrum to the approximate Rayleigh resolution criterion for a data set of approximately 20~d in length, but preserves the amplitude content from the original high-resolution and oversampled Fourier transform. This step is necessary to produce a figure with structure that can be resolved by the human eye. Also, it has the added advantage of homogenising the amplitude spectra in the SC ensemble irrespective of data set length. The modified amplitude spectra, shown between $0 < \nu < 90$~d$^{-1}$, are stacked in order of increasing effective temperature using KIC values and are shown in Fig.~\ref{figure: FT contour 1}. In this figure, solid red lines are used to indicate a moving average of the observables $\nu_{\rm low}$ and $\nu_{\rm high}$ along the star number ordinate axis. The stacked amplitude spectra show an increase in the frequencies of pulsation modes for increasing $T_{\rm eff}$, but also show an increase in the {\it range} of observed pulsation mode frequencies for hotter \dsct stars. Therefore, not only does the frequency of maximum amplitude, $\nu_{\rm max}$, increase with increasing $T_{\rm eff}$, but so do the observables $\nu_{\rm low}$ and $\nu_{\rm high}$.

\begin{figure}
\centering
\includegraphics[width=\columnwidth]{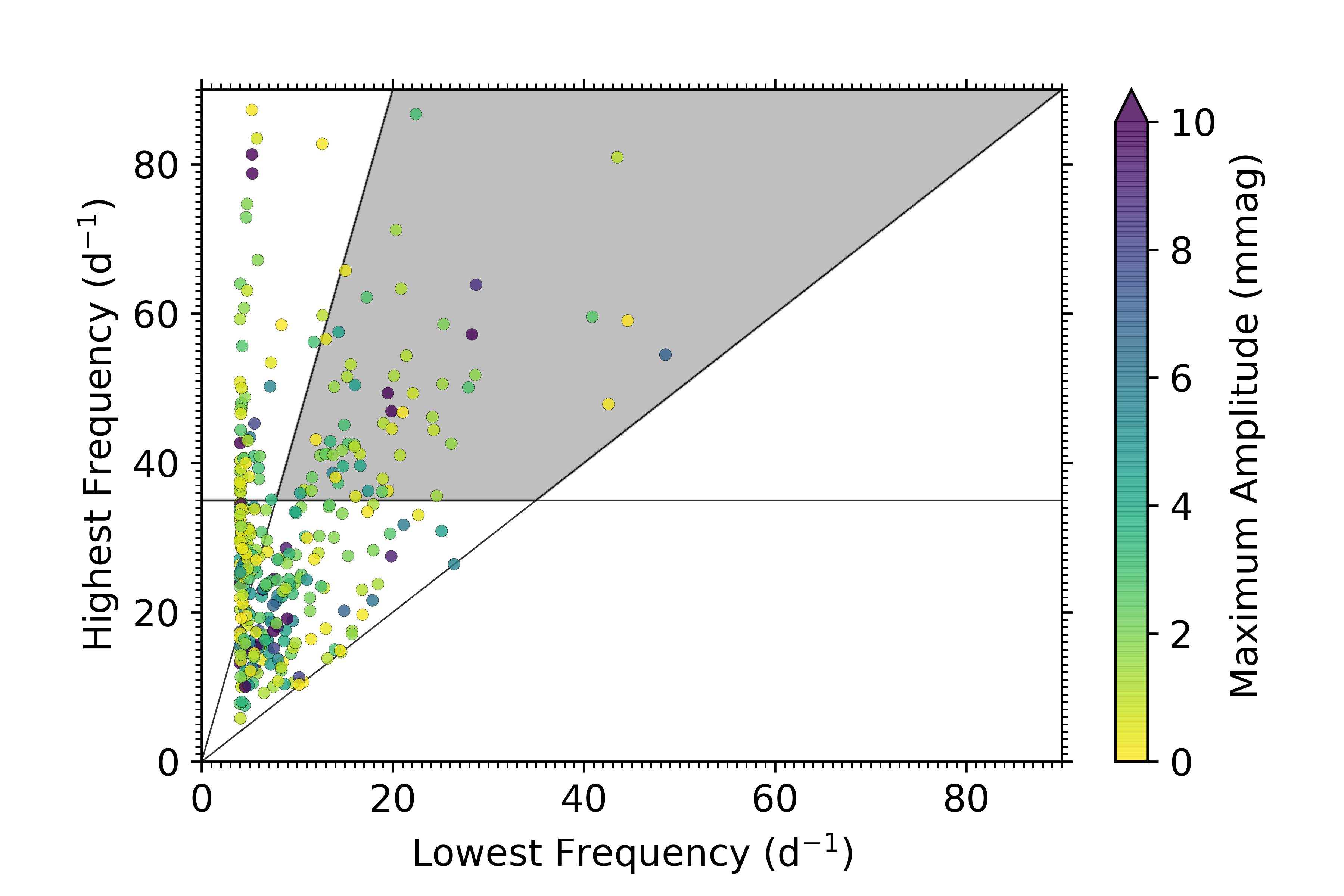}
\caption{The distribution of lowest and highest frequencies of peaks with amplitudes greater than ten times the mean amplitude in the amplitude spectrum above $\nu \geq 4$~d$^{-1}$ for the ensemble of SC \dsct stars. Each star has been colour-coded by the amplitude of the highest-amplitude peak in the amplitude spectrum and the grey region denotes the subsample of young \dsct stars based on $\nu_{\rm high} > 35$~d$^{-1}$ and a ratio of ($\nu_{\rm high}/\nu_{\rm low}) < 4.5$. Since $\nu_{\rm low}$ and $\nu_{\rm high}$ are extracted in the frequency range of $4 \leq \nu \leq 98$~d$^{-1}$, they can be significantly smaller or larger than the corresponding $\nu_{\rm max}$ value for each star (c.f., section~\ref{section: characterising}).}
\label{figure: FT low-high}
\end{figure}

The distribution of the $\nu_{\rm high}$ and $\nu_{\rm low}$ values for each star in the SC ensemble is shown in Fig.~\ref {figure: FT low-high}, with each star shown as a filled circle colour-coded by the amplitude of the dominant pulsation mode. Using the same criteria as \citet{Michel2017b}, a subsample of young \dsct stars was created by including the \dsct stars with $\nu_{\rm high} \geq 35$~d$^{-1}$ ($\gtrsim 400~\mu$Hz) and the ratio of the highest and lowest frequency in each star satisfying $(\nu_{\rm high} / \nu_{\rm low}) < 4.5$. These selection criteria are shown graphically by the individual solid black lines in Fig.~\ref{figure: FT low-high}, with the enclosed grey region indicating the subsample of young \dsct stars. This subgroup contains {60} \dsct stars, which is noticeably fewer than the $\sim$250 CoRoT stars presented by \citet{Michel2017b}. To include more young stars using the SC \Kepler data, the $\nu_{\rm high}$ restriction would have to be relaxed substantially below 35~d$^{-1}$, but this boundary is motivated by theoretical models of pulsation mode frequencies in young \dsct stars being above 35~d$^{-1}$ and changing this criterion removes the compatibility with the analysis by \citet{Michel2017b}.

\begin{figure*}
\centering
\includegraphics[width=\textwidth]{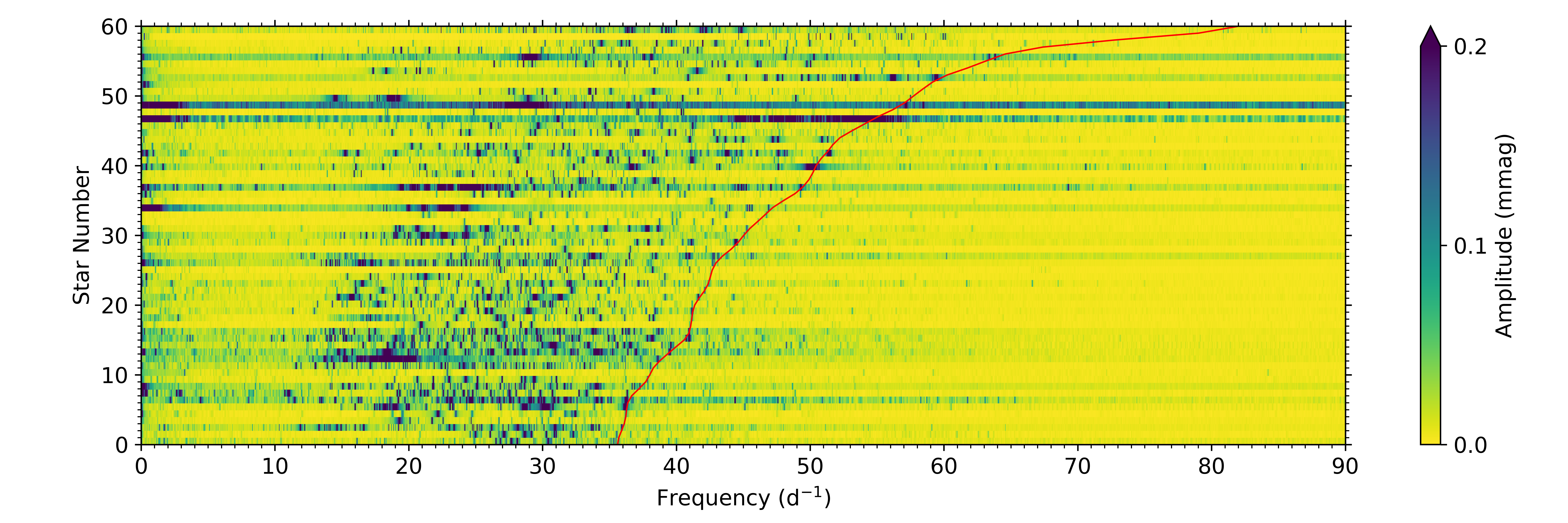}
\includegraphics[width=\textwidth]{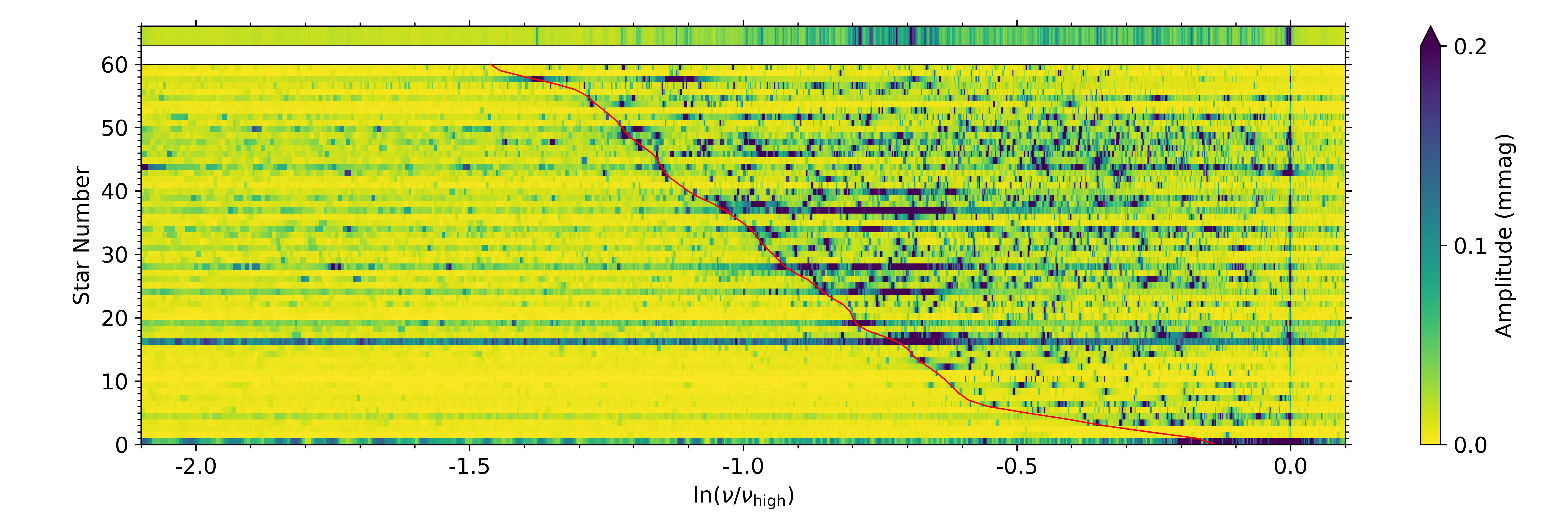}
\caption{The top panel shows the subsample of young \dsct stars from the SC ensemble, which were chosen as the stars that satisfied $\nu_{\rm high} > 35$~d$^{-1}$ and a ratio of ($\nu_{\rm high} / \nu_{\rm low}) < 4.5$ similar to \citet{Michel2017b}. The solid red line indicates a moving average of the $\nu_{\rm high}$ values along the star number ordinate axis. The bottom panel is the same subgroup of stars as the top panel, but with each star's amplitude spectrum having been normalised by its $\nu_{\rm high}$ value and shown on a logarithmic abscissa scale. The solid red line indicates a moving average of the $\nu_{\rm low}$ values along the star number ordinate axis. In the bottom panel, the three rows above the white space show an average of the 60 normalised amplitude spectra. The colour scale in both panels has been purposefully chosen to have an upper limit of 0.2~mmag to reveal low-amplitude peaks, which is similar to the arbitrary upper value of 0.2~parts-per-thousand (ppt) chosen by \citet{Michel2017b}.}
\label{figure: FT contour 2}
\end{figure*}

The stacked amplitude spectra of this subgroup of young \dsct stars using SC \Kepler data are plotted in the top panel of Fig.~\ref{figure: FT contour 2} in order of ascending values of $\nu_{\rm high}$. The high-frequency ridges observed in the amplitude spectra of young \dsct stars studied by \citet{Michel2017b} were shown to be consistent with axisymmetric island modes using theoretical models of fast rotating stars \citep{Reese2009a}. To examine any similar ridges in the amplitude spectra using SC \Kepler data, the amplitude spectra were normalised by $\nu_{\rm high}$ and are plotted on a logarithmic frequency scale in the bottom panel of Fig.~\ref{figure: FT contour 2}, in which the rows above the white space show an average amplitude spectrum for the {60} \dsct stars in the panel. The ridges and regularities discussed by \citet{Michel2017b} are evident in a fraction, but not all, of the amplitude spectra shown in Fig.~\ref{figure: FT contour 2}. Furthermore, the ridge-like structure is not always near the value of $\nu_{\rm high}$ in a star's amplitude spectrum. For example, the last star in the top panel of Fig.~\ref{figure: FT contour 2} has several ridges separated by approximately 3~d$^{-1}$ in its amplitude spectrum between $35 \leq \nu \leq 45$~d$^{-1}$, which could correspond to the large frequency separation in this star, but the value of $\nu_{\rm high} \simeq 80$~d$^{-1}$ using the $S/N \geq 10$ amplitude criterion described previously clearly corresponds to higher frequencies that are harmonics and combination frequencies of these ridges and not intrinsic pulsation modes. 

Other stars similarly have ridges in their amplitude spectra that do not appear near their $\nu_{\rm high}$ values, which explains the lack of regularity in the average amplitude spectrum shown in the rows above the white space in the bottom panel in Fig.~\ref{figure: FT contour 2}. Therefore, this method of normalising amplitude spectrum by $\nu_{\rm high}$ is strongly dependent on the selection criteria used in determining $\nu_{\rm high}$ and is not consistent for all stars. This is easily understood since all peaks in a star's amplitude spectra above $S/N \geq 10$ in amplitude were included in the determination of $\nu_{\rm low}$ and $\nu_{\rm high}$. However, it is possible for a \dsct star to have harmonics and combination frequencies that have amplitudes larger than the chosen amplitude criterion, so a larger $\nu_{\rm high}$ value can be used for such a star which corresponds to a harmonic or a combination frequency rather than a pulsation mode frequency.

The frequency spacing between consecutive radial order p~modes in a \dsct star is of order a few d$^{-1}$, but the specific value depends on the stellar parameters and the radial orders of the pulsation modes \citep{Breger2009a, GH2015, VanReeth2018a**}. Furthermore, although p~modes in \dsct stars are typically low-radial order, pulsation modes of higher radial order can be considered in the transition region where the asymptotic relation for p~modes applies, such that p~modes are equally-spaced in frequency. The mode-dependent transition in the validity of the asymptotic relation for p~modes would break the expectation of finding equally-spaced ridges in the logarithmic plot shown in the bottom panel of Fig.~\ref{figure: FT contour 2}, but the amplitude spectra in linear unnormalised frequencies would remain unaffected for high radial orders. This is certainly true for the stars in Fig.~\ref{figure: Teff-logg} that lie hotter than the blue edge of the classical instability strip and are inferred to be pulsating in p~modes with radial orders of $n \geq 6$.

Therefore, the stars identified as young \dsct stars in Fig.~\ref{figure: FT contour 2}, which were selected because they have high pulsation mode frequencies ($\nu_{\rm high} \geq 35$~d$^{-1}$) corresponding to young stars and inferred to be close to the ZAMS based on predictions from theoretical models \citep{Michel2017b}, represent a subgroup of \dsct stars for which mode identification is simplified because of the presence of regularities in their amplitude spectra. Further observational and theoretical study for the stars showing regularities in their amplitude spectra is required, as they provide insight of mode excitation and interior physics such as rotation and mixing in \dsct stars where mode identification is possible, {c.f., \citet{VanReeth2018a**}}.


\section{Discussion and Conclusions}
\label{section: conclusions}

Two ensembles of \dsct stars using LC and SC \Kepler data consisting of {963} and {334} stars, respectively, were compiled and used to characterise the ensemble pulsational properties of intermediate-mass A and F stars. The LC \Kepler sampling frequency of 48.9~d$^{-1}$ introduces a bias towards extracting low-frequency pulsation modes in an iterative pre-whitening procedure, with higher frequencies being suppressed in amplitude. The amplitude visibility function given in Eqn.~(\ref{equation: amp visibility}) explains the dearth of \dsct stars with pulsation mode frequencies above $\nu \geq 40$~d$^{-1}$ in LC \Kepler data. The amplitude visibility function is negligible in the SC ensemble since the SC sampling frequency is 1476.9~d$^{-1}$; the limitation of these data is that fewer \dsct stars were observed in SC than in LC, and typically not for time spans longer than 30~d which has a poor frequency resolution in comparison. 

The distributions of \dsct stars in $T_{\rm eff} - \log\,g$ diagrams shown in Fig.~\ref{figure: Teff-logg} demonstrate that the theoretical edges of the classical instability strip for modes of radial orders between $1 \leq n \leq 6$ calculated by \citet{Dupret2005} are mostly consistent with observations of \dsct stars from the \Kepler Space Telescope. However, a minority of \dsct stars, including some with stellar parameters obtained using high-resolution spectroscopy by \citet{Tkachenko2012a, Tkachenko2013a} and \citet{Niemczura2015, Niemczura2017a}, are hotter than the blue edge and/or cooler than the red edge of the classical instability strip depending on the source of stellar parameters (e.g., \citealt{Brown2011, Huber2014}). Furthermore, the boundaries of the classical instability strip calculated by \citet{Dupret2005} were calibrated using ground-observations of \dsct stars, which typically have high levels of noise in their amplitude spectra, thus are limited to high pulsation mode amplitudes. The analysis of a large number of \dsct stars in this work supports the requirement for a mass-dependent value of the $\alpha_{\rm MLT}$ parameter in theoretical models to reproduce {\it all} \dsct stars in a $T_{\rm eff} - \log\,g$ diagram. This is particularly necessary to explain the hot \dsct stars that are beyond the blue edge of the $n = 6$ radial mode using $\alpha_{\rm MLT} = 1.8$ \citep{Dupret2004, Dupret2005}.
 
The distributions of maximum pulsation amplitude for the LC and SC ensembles were shown to be consistent with each other, as shown in Fig.~\ref{figure: Amax}, with typical pulsation mode amplitudes between 0.5~mmag and 10~mmag found for most \dsct stars. Similar high-amplitude tails in the amplitude distributions of the LC and SC data are also evident, which are caused by the presence of HADS stars in the \Kepler mission data. These stars have been previously demonstrated to be rare \citep{Lee2008, Balona2016b, Bowman_BOOK}, a finding that is supported by the scarcity of these stars in both LC and SC ensembles in this work. It remains unclear if HADS stars are physically distinct from their low-amplitude counterparts \citep{Breger2000b, Balona2016b, Bowman_PhD, Bowman_BOOK}. Further work is needed to address this.

Despite the lack of amplitude suppression, few \dsct stars exist in the \Kepler data set with pulsation mode frequencies above $60$~d$^{-1}$, with no \dsct stars having their dominant pulsation mode above 60~d$^{-1}$ as shown in Fig.~\ref{figure: freq}. This is in agreement with previous analyses of \dsct stars using subsets of \Kepler data \citep{Balona2011g} if such studies had been corrected for the amplitude visibility function (Eqn.~\ref{equation: amp visibility}). The dearth of high frequency \dsct stars observed by \Kepler can be explained by the lack of hot ZAMS stars, with the TAMS being better sampled in comparison \citep{Tkachenko2012a, Tkachenko2013a, Niemczura2015, Niemczura2017a}. The correlations between the pulsation mode frequencies, effective temperature and evolutionary stage ($\log\,g$ by proxy) were investigated and shown in Fig.~\ref{figure: freq}. The ensembles were separated into ZAMS ($\log\,g \geq 4.0$), MAMS ($3.5 \leq \log\,g < 4.0$) and TAMS ($\log\,g < 3.5$) subgroups using the KIC values, with separate linear regressions between the frequency of the highest-amplitude pulsation mode and $T_{\rm eff}$ for each $\log\,g$ subgroup demonstrating a stronger correlation for ZAMS stars in the SC data. For a ZAMS \dsct star near the blue edge of the classical instability strip, a high effective temperature facilitates the excitation of higher overtone p~modes and higher observed pulsation mode frequencies because the $\kappa$~mechanism operating in the He~{\sc ii} ionisation zone is closer to the surface of the star \citep{Pamyat1999a, Pamyat2000a, Dupret2004, Dupret2005}. This relationship was also found in ground-based observations of \dsct stars \citep{Breger1975, Breger2000b, Rod2001}.

Following a similar methodology to that employed by \citet{Michel2017b} who classified and studied approximately 250 young \dsct stars observed by CoRoT, regularities were searched for in the amplitude spectra of {60} young \dsct stars identified in the SC ensemble of \Kepler observations using the same requirement of high-frequency ($\nu \geq 35$~d$^{-1}$) pulsation modes in their amplitude spectra. Similar ridges consistent with consecutive radial order p~modes can be seen at high frequency in the amplitude spectra of some, but not all, of these young \dsct stars in the SC ensemble. This can partially be explained by the amplitude significance criterion used to determine the observables $\nu_{\rm low}$ and $\nu_{\rm high}$, specifically how harmonics and combination frequencies need to be identified when searching for the frequency separation between radial order p modes. However, a fraction of the stars shown in Fig.~\ref{figure: FT contour 2} do show frequency ridges with spacings of order a few d$^{-1}$, which are consistent with axisymmetric island modes predicted by theoretical models of fast rotating stars \citep{Reese2009a, Michel2017b}. 

The lack of a complete theory for pulsation mode excitation, specifically the non-linear effects that determine the amplitudes of pulsation modes, means we are unable to realistically predict the amplitudes of pulsation modes using current theoretical models. To test if one expects significant regularities caused by the separation between consecutive radial modes in the amplitude spectra of \dsct stars, \citet{Reese2017a} investigated multiplying the intrinsic amplitudes from theoretical amplitude spectra by random numbers uniformly drawn from between 1 and 100 on a logarithmic scale. This proxy for a non-linear mode saturation mechanism, combined with fast rotation, pulsation mode visibility and observing a star at an unknown inclination angle led \citet{Reese2017a} to conclude that finding such regularities in \dsct stars is unlikely. The exceptions to this were found in so-called favourable scenarios where $2\Omega$ coincided with $\Delta\nu$ or $\Delta\nu/2$ corresponding to 30 and 70~per~cent of the break-up velocity \citep{Reese2017a}. The lack of significant regularities in the majority of the amplitude spectra of young \dsct stars presented in this work is consistent with the conclusions presented by \citet{Reese2017a}. On the other, those \dsct stars with regularities consistent with consecutive radial order p~modes may provide valuable constraints of rotation and inclination of \dsct stars based on the discussion by \citet{Reese2017a}, which are typically difficult parameters to determine for \dsct stars.

Other added complications when studying the amplitude spectra of \dsct stars are non-linearity in the form of mode coupling between pulsation modes, and non-linearity in the form of harmonics and combination frequencies \citep{Breger2014, Bowman_BOOK}, which create pseudo-regularities and dense forest-like amplitude spectra. It is also important to note that the pulsation modes in \dsct stars can potentially cover a large range of radial orders such that the expected frequency separation between consecutive radial order p~modes in a star is not a constant. This can be understood from the transition from moderate to high values of $n$ coinciding with the transition for which the asymptotic approximation applies. This produces a non-constant value for the frequency separation across several radial orders in \dsct stars further complicating the task or finding regularity in a star's amplitude spectrum (see e.g., \citealt{Breger2009a}). All the discussed effects act towards blurring the mean of the normalised amplitude spectrum for an ensemble of \dsct stars, which explains the lack of regularity in the average amplitude spectrum shown in the rows above the white space in the bottom panel of Fig.~\ref{figure: FT contour 2}. 

Many \dsct stars are observed to have low-frequency peaks in their amplitude spectra \citep{Balona2014a, Balona2015e}, which can be identified as combination frequencies or independent g-mode pulsations (see e.g., \citealt{Saio2018a}). The full scientific potential of these hybrid stars is yet to be established, with direct measurements of rotation and angular momentum transport needed in A stars needed to fill the gap between B and F stars and constrain theoretical models \citep{VanReeth2016a, Aerts2017b}. Asteroseismic modelling of the most-promising \dsct stars, for which mode identification is possible, using the stellar evolution code {\sc MESA} \citep{Paxton2011, Paxton2013, Paxton2015} and the pulsation code {\sc GYRE} \citep{Townsend2013b}, will constrain theoretical models of interior physics in these stars (see, e.g., \citealt{VanReeth2018a**}). The ensemble of {963} \dsct stars observed continuously by the \Kepler Space Telescope with an unprecedented photometric precision is the best data sets for studying the interior physics of A stars and will remain of great use for many years to come.


\section*{Acknowledgements}

DMB would like to thank the \Kepler science team for providing such excellent data, and Prof. Conny Aerts, Dr. P{\' e}ter P{\' a}pics and Dr. Timothy Van Reeth for useful discussions. Funding for the \Kepler mission is provided by NASA's Science Mission Directorate. The \Kepler data presented in this paper were obtained from the Mikulski Archive for Space Telescopes (MAST). Support for MAST for non-HST data is provided by the NASA Office of Space Science via grant NNX09AF08G and by other grants and contracts. The research leading to these results has received funding from the European Research Council (ERC) under the European Union's Horizon 2020 research and innovation programme (grant agreement N$^{\rm o}$670519: MAMSIE). This research has made use of the SIMBAD database, operated at CDS, Strasbourg, France; the SAO/NASA Astrophysics Data System; and the VizieR catalogue access tool, CDS, Strasbourg, France. We thank the referee for their comments that improved the manuscript.


\bibliographystyle{mnras}
\bibliography{/Users/Dom/Documents/Research/Bibliography/master_bib.bib}

\begin{thebibliography}{}
\makeatletter
\relax
\def\mn@urlcharsother{\let\do\@makeother \do\$\do\&\do\#\do\^\do\_\do\%\do\~}
\def\mn@doi{\begingroup\mn@urlcharsother \@ifnextchar [ {\mn@doi@}
  {\mn@doi@[]}}
\def\mn@doi@[#1]#2{\def\@tempa{#1}\ifx\@tempa\@empty \href
  {http://dx.doi.org/#2} {doi:#2}\else \href {http://dx.doi.org/#2} {#1}\fi
  \endgroup}
\def\mn@eprint#1#2{\mn@eprint@#1:#2::\@nil}
\def\mn@eprint@arXiv#1{\href {http://arxiv.org/abs/#1} {{\tt arXiv:#1}}}
\def\mn@eprint@dblp#1{\href {http://dblp.uni-trier.de/rec/bibtex/#1.xml}
  {dblp:#1}}
\def\mn@eprint@#1:#2:#3:#4\@nil{\def\@tempa {#1}\def\@tempb {#2}\def\@tempc
  {#3}\ifx \@tempc \@empty \let \@tempc \@tempb \let \@tempb \@tempa \fi \ifx
  \@tempb \@empty \def\@tempb {arXiv}\fi \@ifundefined
  {mn@eprint@\@tempb}{\@tempb:\@tempc}{\expandafter \expandafter \csname
  mn@eprint@\@tempb\endcsname \expandafter{\@tempc}}}

\bibitem[\protect\citeauthoryear{{Abdul-Masih} et~al.,}{{Abdul-Masih}
  et~al.}{2016}]{Abdul-Masih2016a}
{Abdul-Masih} M.,  et~al., 2016, \mn@doi [\aj] {10.3847/0004-6256/151/4/101},
  \href {http://adsabs.harvard.edu/abs/2016AJ....151..101A} {151, 101}

\bibitem[\protect\citeauthoryear{{Aerts} \& {Rogers}}{{Aerts} \&
  {Rogers}}{2015}]{Aerts2015c}
{Aerts} C.,  {Rogers} T.~M.,  2015, \mn@doi [\apjl]
  {10.1088/2041-8205/806/2/L33}, \href
  {http://adsabs.harvard.edu/abs/2015ApJ...806L..33A} {806, L33}

\bibitem[\protect\citeauthoryear{{Aerts}, {Christensen-Dalsgaard}  \&
  {Kurtz}}{{Aerts} et~al.}{2010}]{ASTERO_BOOK}
{Aerts} C.,  {Christensen-Dalsgaard} J.,   {Kurtz} D.~W.,  2010,
  Asteroseismology.
Springer

\bibitem[\protect\citeauthoryear{{Aerts} et~al.,}{{Aerts}
  et~al.}{2017a}]{Aerts2017a}
{Aerts} C.,  et~al., 2017a, \mn@doi [\aap] {10.1051/0004-6361/201730571}, \href
  {http://adsabs.harvard.edu/abs/2017A%26A...602A..32A} {602, A32}

\bibitem[\protect\citeauthoryear{{Aerts}, {Van Reeth}  \& {Tkachenko}}{{Aerts}
  et~al.}{2017b}]{Aerts2017b}
{Aerts} C.,  {Van Reeth} T.,   {Tkachenko} A.,  2017b, \mn@doi [\apjl]
  {10.3847/2041-8213/aa8a62}, \href
  {http://adsabs.harvard.edu/abs/2017ApJ...847L...7A} {847, L7}

\bibitem[\protect\citeauthoryear{{Aerts} et~al.,}{{Aerts}
  et~al.}{2018}]{Aerts2018a*}
{Aerts} C.,  et~al., 2018, preprint, \href
  {http://adsabs.harvard.edu/abs/2018arXiv180200621A} {} (\mn@eprint {arXiv}
  {1802.00621})

\bibitem[\protect\citeauthoryear{{Antoci} et~al.,}{{Antoci}
  et~al.}{2014}]{Antoci2014b}
{Antoci} V.,  et~al., 2014, \mn@doi [\apj] {10.1088/0004-637X/796/2/118}, \href
  {http://adsabs.harvard.edu/abs/2014ApJ...796..118A} {796, 118}

\bibitem[\protect\citeauthoryear{{Auvergne} et~al.,}{{Auvergne}
  et~al.}{2009}]{Auvergne2009}
{Auvergne} M.,  et~al., 2009, \mn@doi [\aap] {10.1051/0004-6361/200810860},
  \href {http://adsabs.harvard.edu/abs/2009A%26A...506..411A} {506, 411}

\bibitem[\protect\citeauthoryear{{Balona}}{{Balona}}{2014}]{Balona2014a}
{Balona} L.~A.,  2014, \mn@doi [\mnras] {10.1093/mnras/stt1981}, \href
  {http://adsabs.harvard.edu/abs/2014MNRAS.437.1476B} {437, 1476}

\bibitem[\protect\citeauthoryear{{Balona}}{{Balona}}{2016}]{Balona2016b}
{Balona} L.~A.,  2016, \mn@doi [\mnras] {10.1093/mnras/stw671}, \href
  {http://adsabs.harvard.edu/abs/2016MNRAS.459.1097B} {459, 1097}

\bibitem[\protect\citeauthoryear{{Balona} \& {Dziembowski}}{{Balona} \&
  {Dziembowski}}{2011}]{Balona2011g}
{Balona} L.~A.,  {Dziembowski} W.~A.,  2011, \mn@doi [\mnras]
  {10.1111/j.1365-2966.2011.19301.x}, \href
  {http://adsabs.harvard.edu/abs/2011MNRAS.417..591B} {417, 591}

\bibitem[\protect\citeauthoryear{{Balona}, {Daszy{\'n}ska-Daszkiewicz}  \&
  {Pamyatnykh}}{{Balona} et~al.}{2015}]{Balona2015e}
{Balona} L.~A.,  {Daszy{\'n}ska-Daszkiewicz} J.,   {Pamyatnykh} A.~A.,  2015,
  \mn@doi [\mnras] {10.1093/mnras/stv1513}, \href
  {http://adsabs.harvard.edu/abs/2015MNRAS.452.3073B} {452, 3073}

\bibitem[\protect\citeauthoryear{{Borucki} et~al.,}{{Borucki}
  et~al.}{2010}]{Borucki2010}
{Borucki} W.~J.,  et~al., 2010, \mn@doi [Science] {10.1126/science.1185402},
  \href {http://adsabs.harvard.edu/abs/2010Sci...327..977B} {327, 977}

\bibitem[\protect\citeauthoryear{{Bouabid}, {Dupret}, {Salmon},
  {Montalb{\'a}n}, {Miglio}  \& {Noels}}{{Bouabid} et~al.}{2013}]{Bouabid2013}
{Bouabid} M.-P.,  {Dupret} M.-A.,  {Salmon} S.,  {Montalb{\'a}n} J.,  {Miglio}
  A.,   {Noels} A.,  2013, \mn@doi [\mnras] {10.1093/mnras/sts517}, \href
  {http://adsabs.harvard.edu/abs/2013MNRAS.429.2500B} {429, 2500}

\bibitem[\protect\citeauthoryear{{Bowman}}{{Bowman}}{2016}]{Bowman_PhD}
{Bowman} D.~M.,  2016, PhD thesis, Jeremiah Horrocks Institute, University of
  Central Lancashire, Preston, UK.

\bibitem[\protect\citeauthoryear{{Bowman}}{{Bowman}}{2017}]{Bowman_BOOK}
{Bowman} D.~M.,  2017, {Amplitude Modulation of Pulsation Modes in Delta Scuti
  Stars}.
Springer, \mn@doi{10.1007/978-3-319-66649-5}

\bibitem[\protect\citeauthoryear{{Bowman} \& {Kurtz}}{{Bowman} \&
  {Kurtz}}{2014}]{Bowman2014}
{Bowman} D.~M.,  {Kurtz} D.~W.,  2014, \mn@doi [\mnras]
  {10.1093/mnras/stu1583}, \href
  {http://adsabs.harvard.edu/abs/2014MNRAS.444.1909B} {444, 1909}

\bibitem[\protect\citeauthoryear{{Bowman}, {Holdsworth}  \& {Kurtz}}{{Bowman}
  et~al.}{2015}]{Bowman2015a}
{Bowman} D.~M.,  {Holdsworth} D.~L.,   {Kurtz} D.~W.,  2015, \mn@doi [\mnras]
  {10.1093/mnras/stv364}, \href
  {http://adsabs.harvard.edu/abs/2015MNRAS.449.1004B} {449, 1004}

\bibitem[\protect\citeauthoryear{{Bowman}, {Kurtz}, {Breger}, {Murphy}  \&
  {Holdsworth}}{{Bowman} et~al.}{2016}]{Bowman2016a}
{Bowman} D.~M.,  {Kurtz} D.~W.,  {Breger} M.,  {Murphy} S.~J.,   {Holdsworth}
  D.~L.,  2016, \mn@doi [\mnras] {10.1093/mnras/stw1153}, \href
  {http://adsabs.harvard.edu/abs/2016MNRAS.460.1970B} {460, 1970}

\bibitem[\protect\citeauthoryear{{Breger}}{{Breger}}{2000a}]{Breger2000a}
{Breger} M.,  2000a, in {Szabados} L.,  {Kurtz} D.,  eds,  Astronomical Society
  of the Pacific Conference Series Vol. 203, IAU Colloq. 176: The Impact of
  Large-Scale Surveys on Pulsating Star Research. pp 421--425

\bibitem[\protect\citeauthoryear{{Breger}}{{Breger}}{2000b}]{Breger2000b}
{Breger} M.,  2000b, in {Breger} M.,  {Montgomery} M.,  eds,  Astronomical
  Society of the Pacific Conference Series Vol. 210, Delta Scuti and Related
  Stars. p.~3

\bibitem[\protect\citeauthoryear{{Breger} \& {Bregman}}{{Breger} \&
  {Bregman}}{1975}]{Breger1975}
{Breger} M.,  {Bregman} J.~N.,  1975, \mn@doi [\apj] {10.1086/153794}, \href
  {http://adsabs.harvard.edu/abs/1975ApJ...200..343B} {200, 343}

\bibitem[\protect\citeauthoryear{{Breger} \& {Montgomery}}{{Breger} \&
  {Montgomery}}{2014}]{Breger2014}
{Breger} M.,  {Montgomery} M.~H.,  2014, \mn@doi [\apj]
  {10.1088/0004-637X/783/2/89}, \href
  {http://adsabs.harvard.edu/abs/2014ApJ...783...89B} {783, 89}

\bibitem[\protect\citeauthoryear{{Breger} et~al.,}{{Breger}
  et~al.}{1993}]{Breger1993b}
{Breger} M.,  et~al., 1993, \aap, \href
  {http://adsabs.harvard.edu/abs/1993A%26A...271..482B} {271, 482}

\bibitem[\protect\citeauthoryear{{Breger}, {Lenz}  \& {Pamyatnykh}}{{Breger}
  et~al.}{2009}]{Breger2009a}
{Breger} M.,  {Lenz} P.,   {Pamyatnykh} A.~A.,  2009, \mn@doi [\mnras]
  {10.1111/j.1365-2966.2008.14330.x}, \href
  {http://adsabs.harvard.edu/abs/2009MNRAS.396..291B} {396, 291}

\bibitem[\protect\citeauthoryear{{Breger} et~al.,}{{Breger}
  et~al.}{2012}]{Breger2012b}
{Breger} M.,  et~al., 2012, \mn@doi [\apj] {10.1088/0004-637X/759/1/62}, \href
  {http://adsabs.harvard.edu/abs/2012ApJ...759...62B} {759, 62}

\bibitem[\protect\citeauthoryear{{Brown}, {Latham}, {Everett}  \&
  {Esquerdo}}{{Brown} et~al.}{2011}]{Brown2011}
{Brown} T.~M.,  {Latham} D.~W.,  {Everett} M.~E.,   {Esquerdo} G.~A.,  2011,
  \mn@doi [\aj] {10.1088/0004-6256/142/4/112}, \href
  {http://adsabs.harvard.edu/abs/2011AJ....142..112B} {142, 112}

\bibitem[\protect\citeauthoryear{{Cantiello}, {Mankovich}, {Bildsten},
  {Christensen-Dalsgaard}  \& {Paxton}}{{Cantiello}
  et~al.}{2014}]{Cantiello2014}
{Cantiello} M.,  {Mankovich} C.,  {Bildsten} L.,  {Christensen-Dalsgaard} J.,
  {Paxton} B.,  2014, \mn@doi [\apj] {10.1088/0004-637X/788/1/93}, \href
  {http://adsabs.harvard.edu/abs/2014ApJ...788...93C} {788, 93}

\bibitem[\protect\citeauthoryear{{Chaplin} \& {Miglio}}{{Chaplin} \&
  {Miglio}}{2013}]{Chaplin2013c}
{Chaplin} W.~J.,  {Miglio} A.,  2013, \mn@doi [\araa]
  {10.1146/annurev-astro-082812-140938}, \href
  {http://adsabs.harvard.edu/abs/2013ARA%26A..51..353C} {51, 353}

\bibitem[\protect\citeauthoryear{{Christensen-Dalsgaard}}{{Christensen-Dalsgaard}}{2000}]{C-D2000}
{Christensen-Dalsgaard} J.,  2000, in {Breger} M.,  {Montgomery} M.,  eds,
  Astronomical Society of the Pacific Conference Series Vol. 210, Delta Scuti
  and Related Stars. p.~187

\bibitem[\protect\citeauthoryear{{Cox}}{{Cox}}{1963}]{Cox1963}
{Cox} J.~P.,  1963, \mn@doi [\apj] {10.1086/147661}, \href
  {http://adsabs.harvard.edu/abs/1963ApJ...138..487C} {138, 487}

\bibitem[\protect\citeauthoryear{{Degroote} et~al.,}{{Degroote}
  et~al.}{2010}]{Degroote2010a}
{Degroote} P.,  et~al., 2010, \mn@doi [\nat] {10.1038/nature08864}, \href
  {http://adsabs.harvard.edu/abs/2010Natur.464..259D} {464, 259}

\bibitem[\protect\citeauthoryear{{Dupret}, {Grigahc{\`e}ne}, {Garrido},
  {Gabriel}  \& {Scuflaire}}{{Dupret} et~al.}{2004}]{Dupret2004}
{Dupret} M.~A.,  {Grigahc{\`e}ne} A.,  {Garrido} R.,  {Gabriel} M.,
  {Scuflaire} R.,  2004, \mn@doi [\aap] {10.1051/0004-6361:20031740}, \href
  {http://adsabs.harvard.edu/abs/2004A%26A...414L..17D} {414, L17}

\bibitem[\protect\citeauthoryear{{Dupret}, {Grigahc{\`e}ne}, {Garrido},
  {Gabriel}  \& {Scuflaire}}{{Dupret} et~al.}{2005}]{Dupret2005}
{Dupret} M.~A.,  {Grigahc{\`e}ne} A.,  {Garrido} R.,  {Gabriel} M.,
  {Scuflaire} R.,  2005, \mn@doi [\aap] {10.1051/0004-6361:20041817}, \href
  {http://adsabs.harvard.edu/abs/2005A$\%$26A...435..927D} {435, 927}

\bibitem[\protect\citeauthoryear{{Eggenberger} et~al.,}{{Eggenberger}
  et~al.}{2017}]{Eggenberger2017a}
{Eggenberger} P.,  et~al., 2017, \mn@doi [\aap] {10.1051/0004-6361/201629459},
  \href {http://adsabs.harvard.edu/abs/2017A%26A...599A..18E} {599, A18}

\bibitem[\protect\citeauthoryear{{Garc{\'{\i}}a Hern{\'a}ndez}
  et~al.,}{{Garc{\'{\i}}a Hern{\'a}ndez} et~al.}{2009}]{GH2009}
{Garc{\'{\i}}a Hern{\'a}ndez} A.,  et~al., 2009, \mn@doi [\aap]
  {10.1051/0004-6361/200911932}, \href
  {http://adsabs.harvard.edu/abs/2009A%26A...506...79G} {506, 79}

\bibitem[\protect\citeauthoryear{{Garc{\'{\i}}a Hern{\'a}ndez}
  et~al.,}{{Garc{\'{\i}}a Hern{\'a}ndez} et~al.}{2013}]{GH2013}
{Garc{\'{\i}}a Hern{\'a}ndez} A.,  et~al., 2013, \mn@doi [\aap]
  {10.1051/0004-6361/201220256}, \href
  {http://adsabs.harvard.edu/abs/2013A%26A...559A..63G} {559, A63}

\bibitem[\protect\citeauthoryear{{Garc{\'{\i}}a Hern{\'a}ndez},
  {Mart{\'{\i}}n-Ruiz}, {Monteiro}, {Su{\'a}rez}, {Reese}, {Pascual-Granado}
  \& {Garrido}}{{Garc{\'{\i}}a Hern{\'a}ndez} et~al.}{2015}]{GH2015}
{Garc{\'{\i}}a Hern{\'a}ndez} A.,  {Mart{\'{\i}}n-Ruiz} S.,  {Monteiro}
  M.~J.~P.~F.~G.,  {Su{\'a}rez} J.~C.,  {Reese} D.~R.,  {Pascual-Granado} J.,
  {Garrido} R.,  2015, \mn@doi [\apjl] {10.1088/2041-8205/811/2/L29}, \href
  {http://adsabs.harvard.edu/abs/2015ApJ...811L..29G} {811, L29}

\bibitem[\protect\citeauthoryear{{Gilliland} et~al.,}{{Gilliland}
  et~al.}{2010}]{Gilliland2010}
{Gilliland} R.~L.,  et~al., 2010, \mn@doi [\pasp] {10.1086/650399}, \href
  {http://adsabs.harvard.edu/abs/2010PASP..122..131G} {122, 131}

\bibitem[\protect\citeauthoryear{{Grigahc{\`e}ne}, {Dupret}, {Gabriel},
  {Garrido}  \& {Scuflaire}}{{Grigahc{\`e}ne} et~al.}{2005}]{Griga2005}
{Grigahc{\`e}ne} A.,  {Dupret} M.-A.,  {Gabriel} M.,  {Garrido} R.,
  {Scuflaire} R.,  2005, \mn@doi [\aap] {10.1051/0004-6361:20041816}, \href
  {http://adsabs.harvard.edu/abs/2005A$\%$26A...434.1055G} {434, 1055}

\bibitem[\protect\citeauthoryear{{Grigahc{\`e}ne} et~al.,}{{Grigahc{\`e}ne}
  et~al.}{2010}]{Griga2010a}
{Grigahc{\`e}ne} A.,  et~al., 2010, \mn@doi [\apjl]
  {10.1088/2041-8205/713/2/L192}, \href
  {http://adsabs.harvard.edu/abs/2010ApJ...713L.192G} {713, L192}

\bibitem[\protect\citeauthoryear{{Handler}}{{Handler}}{2009}]{Handler2009c}
{Handler} G.,  2009, \mn@doi [\mnras] {10.1111/j.1365-2966.2009.15005.x}, \href
  {http://adsabs.harvard.edu/abs/2009MNRAS.398.1339H} {398, 1339}

\bibitem[\protect\citeauthoryear{{Hekker} \& {Christensen-Dalsgaard}}{{Hekker}
  \& {Christensen-Dalsgaard}}{2017}]{Hekker2017a}
{Hekker} S.,  {Christensen-Dalsgaard} J.,  2017, \mn@doi [\aapr]
  {10.1007/s00159-017-0101-x}, \href
  {http://adsabs.harvard.edu/abs/2017A%26ARv..25....1H} {25, 1}

\bibitem[\protect\citeauthoryear{{Holdsworth} et~al.,}{{Holdsworth}
  et~al.}{2014}]{Holdsworth2014c}
{Holdsworth} D.~L.,  et~al., 2014, \mn@doi [\mnras] {10.1093/mnras/stu094},
  \href {http://adsabs.harvard.edu/abs/2014MNRAS.439.2078H} {439, 2078}

\bibitem[\protect\citeauthoryear{{Holdsworth}, {Saio}, {Bowman}, {Kurtz},
  {Sefako}, {Joyce}, {Lambert}  \& {Smalley}}{{Holdsworth}
  et~al.}{2018}]{Holdsworth2018b*}
{Holdsworth} D.~L.,  {Saio} H.,  {Bowman} D.~M.,  {Kurtz} D.~W.,  {Sefako}
  R.~R.,  {Joyce} M.,  {Lambert} T.,   {Smalley} B.,  2018, \mn@doi [\mnras]
  {10.1093/mnras/sty248}, \href
  {http://adsabs.harvard.edu/abs/2018MNRAS.tmp..244H} {}

\bibitem[\protect\citeauthoryear{{Houdek}}{{Houdek}}{2000}]{Houdek2000}
{Houdek} G.,  2000, in {Breger} M.,  {Montgomery} M.,  eds,  Astronomical
  Society of the Pacific Conference Series Vol. 210, Delta Scuti and Related
  Stars. p.~454

\bibitem[\protect\citeauthoryear{{Houdek} \& {Dupret}}{{Houdek} \&
  {Dupret}}{2015}]{Houdek2015}
{Houdek} G.,  {Dupret} M.-A.,  2015, \mn@doi [Living Reviews in Solar Physics]
  {10.1007/lrsp-2015-8}, \href
  {http://adsabs.harvard.edu/abs/2015LRSP...12....8H} {12}

\bibitem[\protect\citeauthoryear{{Huber} et~al.,}{{Huber}
  et~al.}{2014}]{Huber2014}
{Huber} D.,  et~al., 2014, \mn@doi [\apjs] {10.1088/0067-0049/211/1/2}, \href
  {http://adsabs.harvard.edu/abs/2014ApJS..211....2H} {211, 2}

\bibitem[\protect\citeauthoryear{{Koch} et~al.,}{{Koch}
  et~al.}{2010}]{Koch2010}
{Koch} D.~G.,  et~al., 2010, \mn@doi [\apjl] {10.1088/2041-8205/713/2/L79},
  \href {http://adsabs.harvard.edu/abs/2010ApJ...713L..79K} {713, L79}

\bibitem[\protect\citeauthoryear{{Kurtz} et~al.,}{{Kurtz}
  et~al.}{2005}]{Kurtz2005b}
{Kurtz} D.~W.,  et~al., 2005, \mn@doi [\mnras]
  {10.1111/j.1365-2966.2005.08807.x}, \href
  {http://adsabs.harvard.edu/abs/2005MNRAS.358..651K} {358, 651}

\bibitem[\protect\citeauthoryear{{Kurtz}, {Saio}, {Takata}, {Shibahashi},
  {Murphy}  \& {Sekii}}{{Kurtz} et~al.}{2014}]{Kurtz2014}
{Kurtz} D.~W.,  {Saio} H.,  {Takata} M.,  {Shibahashi} H.,  {Murphy} S.~J.,
  {Sekii} T.,  2014, \mn@doi [\mnras] {10.1093/mnras/stu1329}, \href
  {http://adsabs.harvard.edu/abs/2014MNRAS.444..102K} {444, 102}

\bibitem[\protect\citeauthoryear{{Kurtz}, {Shibahashi}, {Murphy}, {Bedding}  \&
  {Bowman}}{{Kurtz} et~al.}{2015}]{Kurtz2015b}
{Kurtz} D.~W.,  {Shibahashi} H.,  {Murphy} S.~J.,  {Bedding} T.~R.,   {Bowman}
  D.~M.,  2015, \mn@doi [\mnras] {10.1093/mnras/stv868}, \href
  {http://adsabs.harvard.edu/abs/2015MNRAS.450.3015K} {450, 3015}

\bibitem[\protect\citeauthoryear{{Lee}, {Kim}, {Shin}, {Lee}  \& {Jin}}{{Lee}
  et~al.}{2008}]{Lee2008}
{Lee} Y.-H.,  {Kim} S.~S.,  {Shin} J.,  {Lee} J.,   {Jin} H.,  2008, \mn@doi
  [\pasj] {10.1093/pasj/60.3.551}, \href
  {http://adsabs.harvard.edu/abs/2008PASJ...60..551L} {60, 551}

\bibitem[\protect\citeauthoryear{{Maeder}}{{Maeder}}{2009}]{Maeder_rotation_BOOK}
{Maeder} A.,  2009, {Physics, Formation and Evolution of Rotating Stars}.
Springer, \mn@doi{10.1007/978-3-540-76949-1}

\bibitem[\protect\citeauthoryear{{McNamara}}{{McNamara}}{2000}]{McNamara2000a}
{McNamara} D.~H.,  2000, in {Breger} M.,  {Montgomery} M.,  eds,  Astronomical
  Society of the Pacific Conference Series Vol. 210, Delta Scuti and Related
  Stars. p.~373

\bibitem[\protect\citeauthoryear{{Meynet}, {Ekstrom}, {Maeder}, {Eggenberger},
  {Saio}, {Chomienne}  \& {Haemmerl{\'e}}}{{Meynet}
  et~al.}{2013}]{Meynet_rotation_BOOK}
{Meynet} G.,  {Ekstrom} S.,  {Maeder} A.,  {Eggenberger} P.,  {Saio} H.,
  {Chomienne} V.,   {Haemmerl{\'e}} L.,  2013, in {Goupil} M.,  {Belkacem} K.,
  {Neiner} C.,  {Ligni{\`e}res} F.,   {Green} J.~J.,  eds,  Lecture Notes in
  Physics, Berlin Springer Verlag Vol. 865, Lecture Notes in Physics, Berlin
  Springer Verlag. p.~3 (\mn@eprint {arXiv} {1301.2487}),
  \mn@doi{10.1007/978-3-642-33380-4_1}

\bibitem[\protect\citeauthoryear{{Michel} et~al.,}{{Michel}
  et~al.}{2017}]{Michel2017b}
{Michel} E.,  et~al., 2017, in European Physical Journal Web of Conferences. p.
  03001 (\mn@eprint {arXiv} {1705.03721}),
  \mn@doi{10.1051/epjconf/201716003001}

\bibitem[\protect\citeauthoryear{{Montgomery} \& {O'Donoghue}}{{Montgomery} \&
  {O'Donoghue}}{1999}]{Montgomery1999}
{Montgomery} M.~H.,  {O'Donoghue} D.,  1999, Delta Scuti Star Newsletter, \href
  {http://adsabs.harvard.edu/abs/1999DSSN...13...28M} {13, 28}

\bibitem[\protect\citeauthoryear{{Murphy}, {Shibahashi}  \& {Kurtz}}{{Murphy}
  et~al.}{2013}]{Murphy2013a}
{Murphy} S.~J.,  {Shibahashi} H.,   {Kurtz} D.~W.,  2013, \mn@doi [\mnras]
  {10.1093/mnras/stt105}, \href
  {http://adsabs.harvard.edu/abs/2013MNRAS.430.2986M} {430, 2986}

\bibitem[\protect\citeauthoryear{{Murphy}, {Bedding}, {Niemczura}, {Kurtz}  \&
  {Smalley}}{{Murphy} et~al.}{2015}]{Murphy2015a}
{Murphy} S.~J.,  {Bedding} T.~R.,  {Niemczura} E.,  {Kurtz} D.~W.,   {Smalley}
  B.,  2015, \mn@doi [\mnras] {10.1093/mnras/stu2749}, \href
  {http://adsabs.harvard.edu/abs/2015MNRAS.447.3948M} {447, 3948}

\bibitem[\protect\citeauthoryear{{Murphy}, {Fossati}, {Bedding}, {Saio},
  {Kurtz}, {Grassitelli}  \& {Wang}}{{Murphy} et~al.}{2016}]{Murphy2016a}
{Murphy} S.~J.,  {Fossati} L.,  {Bedding} T.~R.,  {Saio} H.,  {Kurtz} D.~W.,
  {Grassitelli} L.,   {Wang} E.~S.,  2016, \mn@doi [\mnras]
  {10.1093/mnras/stw705}, \href
  {http://adsabs.harvard.edu/abs/2016MNRAS.459.1201M} {459, 1201}

\bibitem[\protect\citeauthoryear{{Niemczura} et~al.,}{{Niemczura}
  et~al.}{2015}]{Niemczura2015}
{Niemczura} E.,  et~al., 2015, \mn@doi [\mnras] {10.1093/mnras/stv528}, \href
  {http://adsabs.harvard.edu/abs/2015MNRAS.450.2764N} {450, 2764}

\bibitem[\protect\citeauthoryear{{Niemczura} et~al.,}{{Niemczura}
  et~al.}{2017}]{Niemczura2017a}
{Niemczura} E.,  et~al., 2017, \mn@doi [\mnras] {10.1093/mnras/stx1256}, \href
  {http://adsabs.harvard.edu/abs/2017MNRAS.470.2870N} {470, 2870}

\bibitem[\protect\citeauthoryear{{Ouazzani}, {Salmon}, {Antoci}, {Bedding},
  {Murphy}  \& {Roxburgh}}{{Ouazzani} et~al.}{2017}]{Ouazzani2017a}
{Ouazzani} R.-M.,  {Salmon} S.~J.~A.~J.,  {Antoci} V.,  {Bedding} T.~R.,
  {Murphy} S.~J.,   {Roxburgh} I.~W.,  2017, \mn@doi [\mnras]
  {10.1093/mnras/stw2717}, \href
  {http://adsabs.harvard.edu/abs/2017MNRAS.465.2294O} {465, 2294}

\bibitem[\protect\citeauthoryear{{Pamyatnykh}}{{Pamyatnykh}}{1999}]{Pamyat1999a}
{Pamyatnykh} A.~A.,  1999, in {Wolf} B.,  {Stahl} O.,   {Fullerton} A.~W.,
  eds,  Lecture Notes in Physics, Berlin Springer Verlag Vol. 523, IAU Colloq.
  169: Variable and Non-spherical Stellar Winds in Luminous Hot Stars. p.~320,
  \mn@doi{10.1007/BFb0106397}

\bibitem[\protect\citeauthoryear{{Pamyatnykh}}{{Pamyatnykh}}{2000}]{Pamyat2000a}
{Pamyatnykh} A.~A.,  2000, in {Breger} M.,  {Montgomery} M.,  eds,
  Astronomical Society of the Pacific Conference Series Vol. 210, Delta Scuti
  and Related Stars. p.~215 (\mn@eprint {} {astro-ph/0005276})

\bibitem[\protect\citeauthoryear{{Papar{\'o}}, {Benk{\H o}}, {Hareter}  \&
  {Guzik}}{{Papar{\'o}} et~al.}{2016}]{Paparo2016b}
{Papar{\'o}} M.,  {Benk{\H o}} J.~M.,  {Hareter} M.,   {Guzik} J.~A.,  2016,
  \mn@doi [\apjs] {10.3847/0067-0049/224/2/41}, \href
  {http://adsabs.harvard.edu/abs/2016ApJS..224...41P} {224, 41}

\bibitem[\protect\citeauthoryear{{P{\'a}pics}}{{P{\'a}pics}}{2012}]{Papics2012b}
{P{\'a}pics} P.~I.,  2012, \mn@doi [Astronomische Nachrichten]
  {10.1002/asna.201211809}, \href
  {http://adsabs.harvard.edu/abs/2012AN....333.1053P} {333, 1053}

\bibitem[\protect\citeauthoryear{{P{\'a}pics}, {Moravveji}, {Aerts},
  {Tkachenko}, {Triana}, {Bloemen}  \& {Southworth}}{{P{\'a}pics}
  et~al.}{2014}]{Papics2014}
{P{\'a}pics} P.~I.,  {Moravveji} E.,  {Aerts} C.,  {Tkachenko} A.,  {Triana}
  S.~A.,  {Bloemen} S.,   {Southworth} J.,  2014, \mn@doi [\aap]
  {10.1051/0004-6361/201424094}, \href
  {http://adsabs.harvard.edu/abs/2014A%26A...570A...8P} {570, A8}

\bibitem[\protect\citeauthoryear{{P{\'a}pics} et~al.,}{{P{\'a}pics}
  et~al.}{2017}]{Papics2017a}
{P{\'a}pics} P.~I.,  et~al., 2017, \mn@doi [\aap]
  {10.1051/0004-6361/201629814}, \href
  {http://adsabs.harvard.edu/abs/2017A%26A...598A..74P} {598, A74}

\bibitem[\protect\citeauthoryear{{Paxton}, {Bildsten}, {Dotter}, {Herwig},
  {Lesaffre}  \& {Timmes}}{{Paxton} et~al.}{2011}]{Paxton2011}
{Paxton} B.,  {Bildsten} L.,  {Dotter} A.,  {Herwig} F.,  {Lesaffre} P.,
  {Timmes} F.,  2011, \mn@doi [\apjs] {10.1088/0067-0049/192/1/3}, \href
  {http://adsabs.harvard.edu/abs/2011ApJS..192....3P} {192, 3}

\bibitem[\protect\citeauthoryear{{Paxton} et~al.,}{{Paxton}
  et~al.}{2013}]{Paxton2013}
{Paxton} B.,  et~al., 2013, \mn@doi [\apjs] {10.1088/0067-0049/208/1/4}, \href
  {http://adsabs.harvard.edu/abs/2013ApJS..208....4P} {208, 4}

\bibitem[\protect\citeauthoryear{{Paxton} et~al.,}{{Paxton}
  et~al.}{2015}]{Paxton2015}
{Paxton} B.,  et~al., 2015, \mn@doi [\apjs] {10.1088/0067-0049/220/1/15}, \href
  {http://adsabs.harvard.edu/abs/2015ApJS..220...15P} {220, 15}

\bibitem[\protect\citeauthoryear{{Pr{\v s}a} et~al.,}{{Pr{\v s}a}
  et~al.}{2011}]{Prsa2011}
{Pr{\v s}a} A.,  et~al., 2011, \mn@doi [\aj] {10.1088/0004-6256/141/3/83},
  \href {http://adsabs.harvard.edu/abs/2011AJ....141...83P} {141, 83}

\bibitem[\protect\citeauthoryear{{Raskin} et~al.,}{{Raskin}
  et~al.}{2011}]{Raskin2011}
{Raskin} G.,  et~al., 2011, \mn@doi [\aap] {10.1051/0004-6361/201015435}, \href
  {http://adsabs.harvard.edu/abs/2011A%26A...526A..69R} {526, A69}

\bibitem[\protect\citeauthoryear{{Reese}, {MacGregor}, {Jackson}, {Skumanich}
  \& {Metcalfe}}{{Reese} et~al.}{2009}]{Reese2009a}
{Reese} D.~R.,  {MacGregor} K.~B.,  {Jackson} S.,  {Skumanich} A.,   {Metcalfe}
  T.~S.,  2009, \mn@doi [\aap] {10.1051/0004-6361/200811510}, \href
  {http://adsabs.harvard.edu/abs/2009A%26A...506..189R} {506, 189}

\bibitem[\protect\citeauthoryear{{Reese}, {Ligni{\`e}res}, {Ballot}, {Dupret},
  {Barban}, {van't Veer-Menneret}  \& {MacGregor}}{{Reese}
  et~al.}{2017}]{Reese2017a}
{Reese} D.~R.,  {Ligni{\`e}res} F.,  {Ballot} J.,  {Dupret} M.-A.,  {Barban}
  C.,  {van't Veer-Menneret} C.,   {MacGregor} K.~B.,  2017, \mn@doi [\aap]
  {10.1051/0004-6361/201321264}, \href
  {http://adsabs.harvard.edu/abs/2017A%26A...601A.130R} {601, A130}

\bibitem[\protect\citeauthoryear{{Rodr{\'{\i}}guez} \&
  {Breger}}{{Rodr{\'{\i}}guez} \& {Breger}}{2001}]{Rod2001}
{Rodr{\'{\i}}guez} E.,  {Breger} M.,  2001, \mn@doi [\aap]
  {10.1051/0004-6361:20000205}, \href
  {http://adsabs.harvard.edu/abs/2001A%26A...366..178R} {366, 178}

\bibitem[\protect\citeauthoryear{{Rogers}}{{Rogers}}{2015}]{Rogers2015}
{Rogers} T.~M.,  2015, \mn@doi [\apjl] {10.1088/2041-8205/815/2/L30}, \href
  {http://adsabs.harvard.edu/abs/2015ApJ...815L..30R} {815, L30}

\bibitem[\protect\citeauthoryear{{Rogers}, {Lin}, {McElwaine}  \&
  {Lau}}{{Rogers} et~al.}{2013}]{Rogers2013b}
{Rogers} T.~M.,  {Lin} D.~N.~C.,  {McElwaine} J.~N.,   {Lau} H.~H.~B.,  2013,
  \mn@doi [\apj] {10.1088/0004-637X/772/1/21}, \href
  {http://adsabs.harvard.edu/abs/2013ApJ...772...21R} {772, 21}

\bibitem[\protect\citeauthoryear{{Saio}, {Kurtz}, {Takata}, {Shibahashi},
  {Murphy}, {Sekii}  \& {Bedding}}{{Saio} et~al.}{2015}]{Saio2015b}
{Saio} H.,  {Kurtz} D.~W.,  {Takata} M.,  {Shibahashi} H.,  {Murphy} S.~J.,
  {Sekii} T.,   {Bedding} T.~R.,  2015, \mn@doi [\mnras]
  {10.1093/mnras/stu2696}, \href
  {http://adsabs.harvard.edu/abs/2015MNRAS.447.3264S} {447, 3264}

\bibitem[\protect\citeauthoryear{{Saio}, {Kurtz}, {Murphy}, {Antoci}  \&
  {Lee}}{{Saio} et~al.}{2018}]{Saio2018a}
{Saio} H.,  {Kurtz} D.~W.,  {Murphy} S.~J.,  {Antoci} V.~L.,   {Lee} U.,  2018,
  \mn@doi [\mnras] {10.1093/mnras/stx2962}, \href
  {http://adsabs.harvard.edu/abs/2018MNRAS.474.2774S} {474, 2774}

\bibitem[\protect\citeauthoryear{{Sim{\'o}n-D{\'{\i}}az}, {Aerts}, {Urbaneja},
  {Camacho}, {Antoci}, {Fredslund Andersen}, {Grundahl}  \&
  {Pall{\'e}}}{{Sim{\'o}n-D{\'{\i}}az} et~al.}{2017}]{Simon-Diaz2018a*}
{Sim{\'o}n-D{\'{\i}}az} S.,  {Aerts} C.,  {Urbaneja} M.~A.,  {Camacho} I.,
  {Antoci} V.,  {Fredslund Andersen} M.,  {Grundahl} F.,   {Pall{\'e}} P.~L.,
  2017, preprint, \href {http://adsabs.harvard.edu/abs/2017arXiv171108994S} {}
  (\mn@eprint {arXiv} {1711.08994})

\bibitem[\protect\citeauthoryear{{Smith} et~al.,}{{Smith}
  et~al.}{2012}]{Smith2012}
{Smith} J.~C.,  et~al., 2012, \mn@doi [\pasp] {10.1086/667697}, \href
  {http://adsabs.harvard.edu/abs/2012PASP..124.1000S} {124, 1000}

\bibitem[\protect\citeauthoryear{{Stumpe} et~al.,}{{Stumpe}
  et~al.}{2012}]{Stumpe2012}
{Stumpe} M.~C.,  et~al., 2012, \mn@doi [\pasp] {10.1086/667698}, \href
  {http://adsabs.harvard.edu/abs/2012PASP..124..985S} {124, 985}

\bibitem[\protect\citeauthoryear{{Tayar} \& {Pinsonneault}}{{Tayar} \&
  {Pinsonneault}}{2013}]{Tayar2013}
{Tayar} J.,  {Pinsonneault} M.~H.,  2013, \mn@doi [\apjl]
  {10.1088/2041-8205/775/1/L1}, \href
  {http://adsabs.harvard.edu/abs/2013ApJ...775L...1T} {775, L1}

\bibitem[\protect\citeauthoryear{{Tkachenko}, {Lehmann}, {Smalley},
  {Debosscher}  \& {Aerts}}{{Tkachenko} et~al.}{2012}]{Tkachenko2012a}
{Tkachenko} A.,  {Lehmann} H.,  {Smalley} B.,  {Debosscher} J.,   {Aerts} C.,
  2012, \mn@doi [\mnras] {10.1111/j.1365-2966.2012.20687.x}, \href
  {http://adsabs.harvard.edu/abs/2012MNRAS.422.2960T} {422, 2960}

\bibitem[\protect\citeauthoryear{{Tkachenko}, {Lehmann}, {Smalley}  \&
  {Uytterhoeven}}{{Tkachenko} et~al.}{2013}]{Tkachenko2013a}
{Tkachenko} A.,  {Lehmann} H.,  {Smalley} B.,   {Uytterhoeven} K.,  2013,
  \mn@doi [\mnras] {10.1093/mnras/stt453}, \href
  {http://adsabs.harvard.edu/abs/2013MNRAS.431.3685T} {431, 3685}

\bibitem[\protect\citeauthoryear{{Tkachenko} et~al.,}{{Tkachenko}
  et~al.}{2014}]{Tkachenko2014a}
{Tkachenko} A.,  et~al., 2014, \mn@doi [\mnras] {10.1093/mnras/stt2421}, \href
  {http://adsabs.harvard.edu/abs/2014MNRAS.438.3093T} {438, 3093}

\bibitem[\protect\citeauthoryear{{Townsend} \& {Teitler}}{{Townsend} \&
  {Teitler}}{2013}]{Townsend2013b}
{Townsend} R.~H.~D.,  {Teitler} S.~A.,  2013, \mn@doi [\mnras]
  {10.1093/mnras/stt1533}, \href
  {http://adsabs.harvard.edu/abs/2013MNRAS.435.3406T} {435, 3406}

\bibitem[\protect\citeauthoryear{{Triana}, {Moravveji}, {P{\'a}pics}, {Aerts},
  {Kawaler}  \& {Christensen-Dalsgaard}}{{Triana} et~al.}{2015}]{Triana2015}
{Triana} S.~A.,  {Moravveji} E.,  {P{\'a}pics} P.~I.,  {Aerts} C.,  {Kawaler}
  S.~D.,   {Christensen-Dalsgaard} J.,  2015, \mn@doi [\apj]
  {10.1088/0004-637X/810/1/16}, \href
  {http://adsabs.harvard.edu/abs/2015ApJ...810...16T} {810, 16}

\bibitem[\protect\citeauthoryear{{Uytterhoeven} et~al.,}{{Uytterhoeven}
  et~al.}{2011}]{Uytterhoeven2011}
{Uytterhoeven} K.,  et~al., 2011, \mn@doi [\aap] {10.1051/0004-6361/201117368},
  \href {http://adsabs.harvard.edu/abs/2011A$\%$26A...534A.125U} {534, A125}

\bibitem[\protect\citeauthoryear{{Van Reeth} et~al.,}{{Van Reeth}
  et~al.}{2015a}]{VanReeth2015b}
{Van Reeth} T.,  et~al., 2015a, \mn@doi [\apjs] {10.1088/0067-0049/218/2/27},
  \href {http://adsabs.harvard.edu/abs/2015ApJS..218...27V} {218, 27}

\bibitem[\protect\citeauthoryear{{Van Reeth} et~al.,}{{Van Reeth}
  et~al.}{2015b}]{VanReeth2015a}
{Van Reeth} T.,  et~al., 2015b, \mn@doi [\aap] {10.1051/0004-6361/201424585},
  \href {http://adsabs.harvard.edu/abs/2015A%26A...574A..17V} {574, A17}

\bibitem[\protect\citeauthoryear{{Van Reeth}, {Tkachenko}  \& {Aerts}}{{Van
  Reeth} et~al.}{2016}]{VanReeth2016a}
{Van Reeth} T.,  {Tkachenko} A.,   {Aerts} C.,  2016, \mn@doi [\aap]
  {10.1051/0004-6361/201628616}, \href
  {http://adsabs.harvard.edu/abs/2016A%26A...593A.120V} {593, A120}

\bibitem[\protect\citeauthoryear{{Van Reeth} et~al.,}{{Van Reeth}
  et~al.}{2018}]{VanReeth2018a**}
{Van Reeth} T.,  et~al., 2018, submitted to \aap

\bibitem[\protect\citeauthoryear{{Xiong}, {Deng}, {Zhang}  \& {Wang}}{{Xiong}
  et~al.}{2016}]{Xiong2016a}
{Xiong} D.~R.,  {Deng} L.,  {Zhang} C.,   {Wang} K.,  2016, \mn@doi [\mnras]
  {10.1093/mnras/stw047}, \href
  {http://adsabs.harvard.edu/abs/2016MNRAS.457.3163X} {457, 3163}

\bibitem[\protect\citeauthoryear{{Zwintz} et~al.,}{{Zwintz}
  et~al.}{2017}]{Zwintz2017a}
{Zwintz} K.,  et~al., 2017, \mn@doi [\aap] {10.1051/0004-6361/201630327}, \href
  {http://adsabs.harvard.edu/abs/2017A%26A...601A.101Z} {601, A101}

\makeatother
\end{thebibliography}


\bsp	
\label{lastpage}
\end{document}